\RequirePackage{fix-cm}
\documentclass[twocolumn]{svjour3}          
\smartqed  
\usepackage{url}
\usepackage{subfig}
\usepackage{caption}
\usepackage{graphicx}
\usepackage{xcolor}
\usepackage{float}
\usepackage[numbers]{natbib}
\begin{document}

\title{Extraction of Layout Entities and Sub-layout Query-based
	Retrieval of Document Images
}


\author{Anukriti Bansal         \and
        Sumantra Dutta Roy 		\and
		Gaurav Harit
}


\institute{A. Bansal \at
              Indian Institute of Technology Delhi \\
				New Delhi, India \\
              \email{anukriti1107@gmail.com}           
           \and
           S. Dutta Roy \at
		   	  Indian Institute of Technology Delhi \\
			  New Delhi, India \\
			  \email{sumantra@ee.iitd.ac.in}
			\and
			G. Harit \at
				Indian Institute of Technology Jodhpur \\
				Rajasthan, India \\
				\email{gharit@iitj.ac.in}
}

\date{Received: date / Accepted: date}

\maketitle

\begin{abstract}
Layouts and sub-layouts constitute an important clue while
searching a document on the basis of its structure, or when
textual content is unknown/irrelevant.  A sub-layout specifies
the arrangement of document entities within a smaller portion of
the document. 
We propose an
efficient graph-based matching algorithm, integrated with
hash-based indexing, to prune a possibly large search space. 
A user can specify a combination of sub-layouts of interest using
sketch-based queries.
The system supports partial matching for unspecified
layout entities.
We handle cases of segmentation pre-processing
errors (for text/non-text blocks) with a symmetry
maximization-based strategy, and accounting for multiple
domain-specific plausible segmentation hypotheses.
We show promising results of our system on
a database of unstructured entities, containing 4776 newspaper images.
\end{abstract}

\section{Introduction}
\label{intro}
The advancement of digitization of document images has increased
the need for an accurate, efficient and user-friendly large-scale
information search and retrieval from the databases and archives.
Document retrieval can be broadly classified into two types: content-based
and layout-based (\cite{Doermann:98}).
Content-based approaches are highly dependent on document
labeling, feature selection and high computation involved for OCR
systems. Another issue with content-based systems is that the
layout information of the document is lost completely which
constitute an important clue while searching for particular type
of information. Besides this, a more fundamental problem with the
content-based system is that the textual content of the document
to be searched for is irrelevant/not known exactly.
Further,
the required information could be present in a small region
(\cite{Aurisicchio:2003} and \cite{Lowe:1999}).
An example includes a portion of a newspaper image with fixed
format and position, e.g., a columnists' article in the leftmost image of
the editorial page in Fig.~\ref{fig:Results_Boolean}, highlighted in orange.
Thus, the sub-layout of blocks/image-entities plays a significant role
in querying for such documents.
A sub-layout can be defined as an
arrangement of blocks within a portion of document/image.
A sub-layout-based method
can enhance existing content-based retrieval by
reducing the set of candidate documents~\cite{Shin:06}.
(A user may seek an article about Bill Gates in the lower half of a
newspaper page, which has a text block followed by
an image on the left and text on the right).

\begin{figure*}[ht]
\centering
\fbox{\includegraphics[width = 2.5cm]{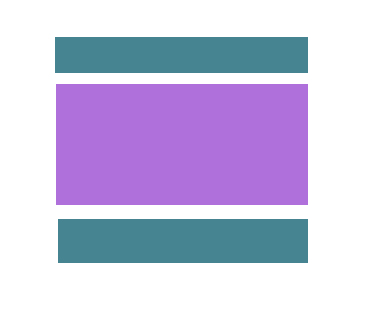}
\label{Thesis_query}}
\fbox{\includegraphics[width = 3cm]{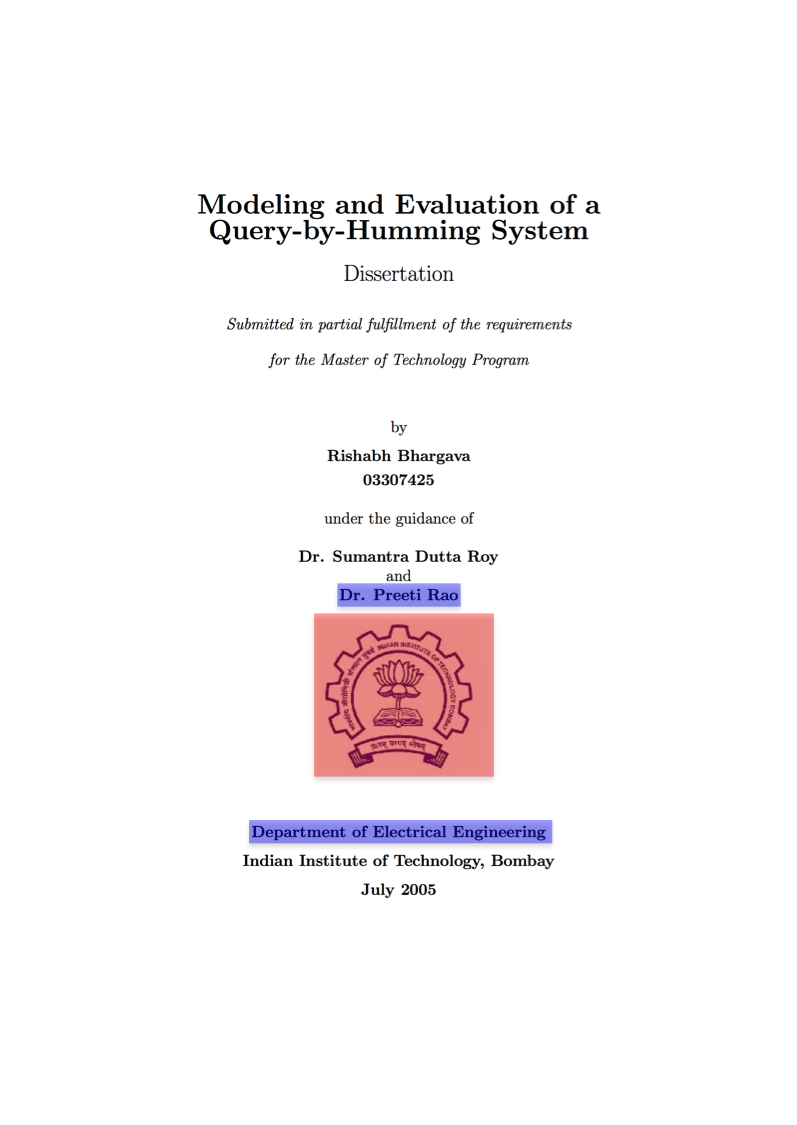}
\label{Thesis1}} \hspace{0.25mm}
\fbox{\includegraphics[width = 3cm]{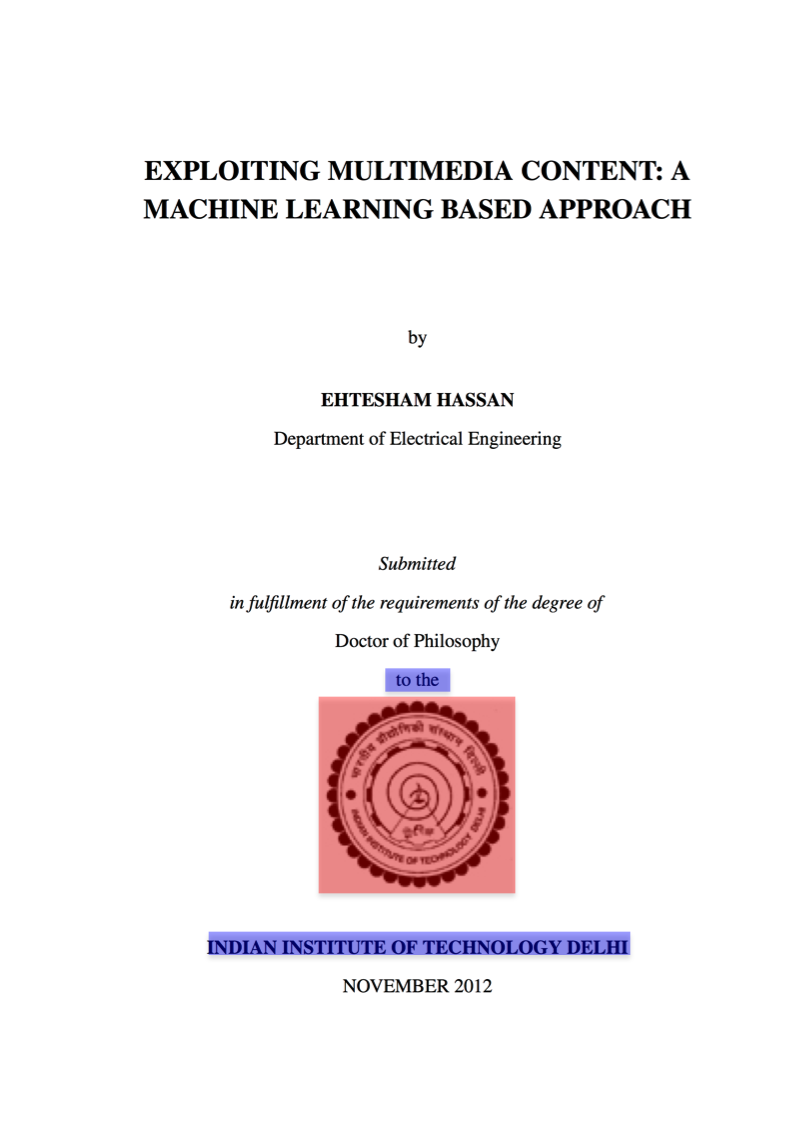}
\label{Thesis2}} \hspace{0.25mm}
\fbox{\includegraphics[width = 3cm]{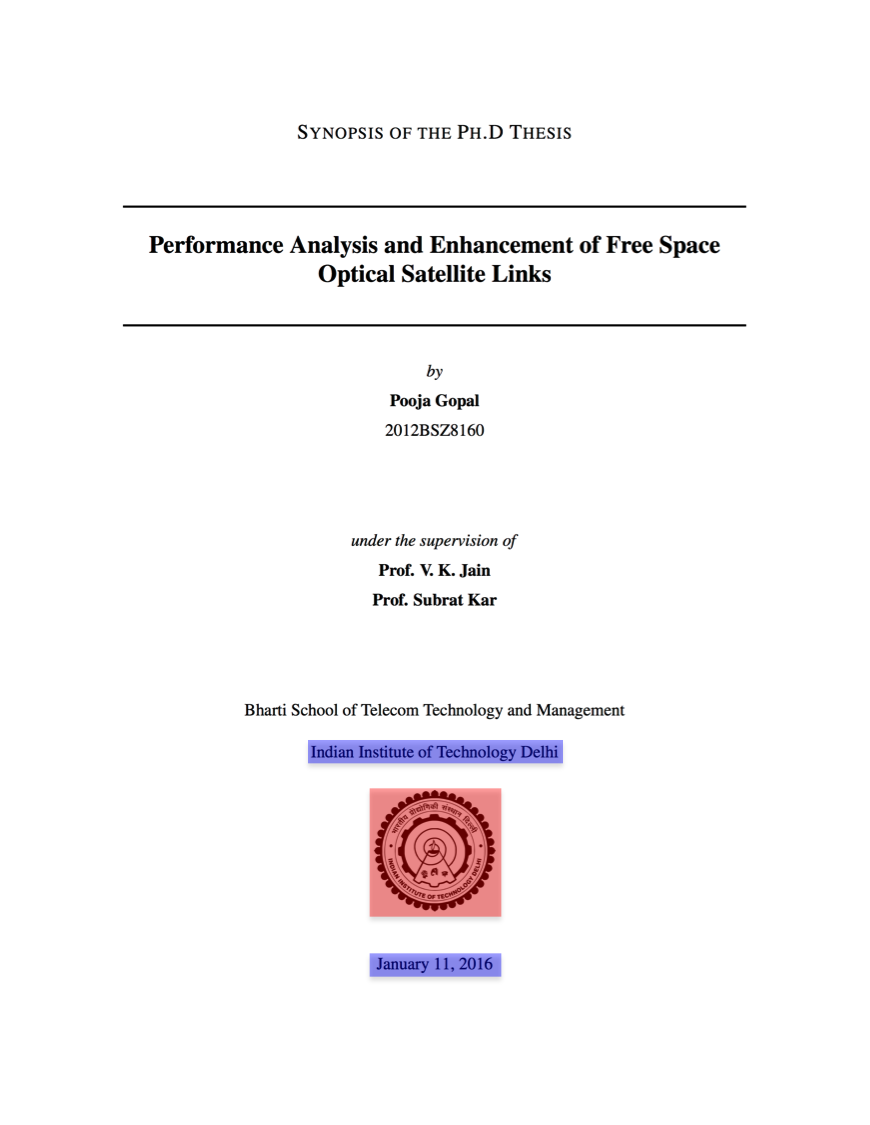}
\label{Thesis3}}
\caption{Application of sub-layout-based retrieval: (a)
	represents the query for image region (shown in pink color)
	sandwiched between text regions (shown in blue color) on both
	the sides and present at the center of the page. (b), (c), (d) are the examples
	of retrieved document images.}
\label{fig:Thesis_example}
\end{figure*}

A sub-layout-based retrieval can be used while designing the
complete page layout of magazines, newspapers and official
documents. A user may want to survey a set of designs which need
some fixed content, and some part of it left to the a designer's
imagination and decision. For example, the cover page of a thesis
has few components fixed, such as the University logo
sandwiched between text. A user can give this sub-layout as a query,
and based on the retrieved results, draw further cues for placing
other components, to design the entire cover page (Figure
\ref{fig:Thesis_example}).

Studies have shown that humans have better memory of images over
text (\cite{Baggett:1979}, \cite{Paivio:1973}). 
Thus, layout information can be used in applications such as
personal filing system, where images are viewed and archived by
an individual, and later retrieved by visual recollection
Neuroscience studies (\cite{Landay:2001}, \cite{Marr:1982})
suggest that sketching is a fundamental way for humans to
conceptualize and render things. This paper uses sketch-based
queries so that a user can specify the desired layout/sub-layout
in a natural and expressive manner.

Delivery of desired information in proper format and right quantity
motivates us for sub-layout-based document image retrieval using
user specified sketch-based queries.
\subsection{Related Work}
\label{sec:Related_Work}
Image retrieval and classification has been an interesting research problem from
last two decades (\cite{Shin:06}, \cite{Marinai:2011},
\cite{Mitra:2000}, \cite{Doermann:98}, \cite{Chen:2007},
\cite{Antani:2002}). 
Layout information can be used for document image classification
as well as retrieval.

%


Hu et al. \cite{Hu:99} compute
the distance between image rows after a segmentation into a grid
of equal-sized cells. Each cell is labeled as text or whitespace
on the basis of its overlapping with the text block. Document
images are then compared using
dynamic programming on a row-based representation of documents.

Hu et al. \cite{Hu:99b} describe methods for document image classification
at the spatial layout level. The elements of spatial layout are
captured by a feature set called interval coding, which encodes
structural layout information of the region using fixed-length
vectors. These features are used in a hidden Markov model-based
page layout classification system. 

Shin et al. \cite{Shin:01} use visually similar features of layout
structure such as the percentage of text and non-text regions,
column structures, relative font sizes, density of the content area
and statistics of features of connected components, for the
classification of document pages using decision tree classifiers
and self-organizing maps.

Kumar et al. \cite{Kumar:2014} use layout and spatial organization of
document image content for image classification and retrieval.
They recursively partition the image and compute histograms of
codewords (SURF descriptors) in each partition. A random forest
classifier is used for classification and retrieval.

The state-of-the-art in camera-based document image retrieval
perhaps comes from Osaka Prefecture University Group, where
Takeda et al. \cite{Takeda:2011} and Nakai et al. \cite{Nakai:06}
have used Locally Likely Arrangement Hashing (LLAH) for real time
document retrieval.
The method uses centroid of word regions as feature points and
calculates geometrically invariant features on
the basis of neighboring feature points.  
The method method works well on camera captured images as
queries, which contains many word images.
This may not work well for queries where word blocks are not large in
number , or those with a large number of non-text 
blocks (images, graphics) \cite{Takeda:2011}. Such a method relies on the actual content,
hence limiting its scope.
A layout-based method may work better in such cases, which
considers the relative arrangement of all types of blocks
(text/non-text). Further, if we extend the LLAH idea from word
blocks to centroids of text/non-text blocks, the structurally
invariant information may not be adequately captured with 
a few layout components. 

van Beusekom et al. \cite{Van:2006} use layout information for document retrieval. A
class of distance measure based on two-step procedure is
introduced. In the first step, the distances between the blocks
of document and query layouts are calculated. 
Various types of distance measures like, Manhattan distance of
corner points, overlapping area of blocks, difference in width
and height, etc., were used to compute distance between every
pair of blocks in given layouts.  
Then, the matching step matches the blocks of query layout to
the blocks of reference layout by minimizing the total distance.
Since the method uses distance measure, it is not invariant to
position and shape of the blocks. Besides, the number of blocks
in case of sub-layouts will be lesser and their position and
aspect ratios can be different, the total distance between two
layouts will be larger and thus, may not match.

\begin{figure}[h]
\centering
\fbox{\includegraphics[width = 3.5cm, height = 4.5cm]{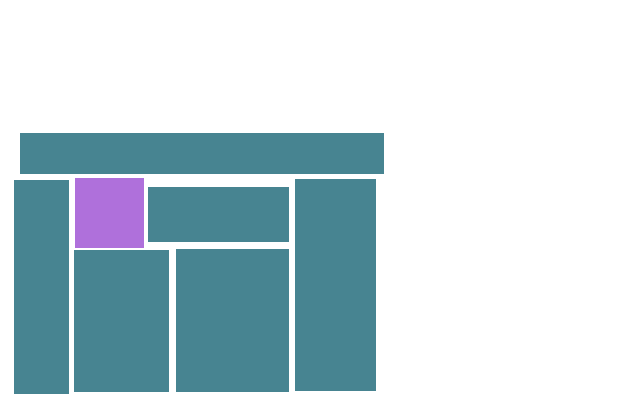}} \hspace{1mm}
\fbox{\includegraphics[width = 3.5cm, height = 4.5cm]{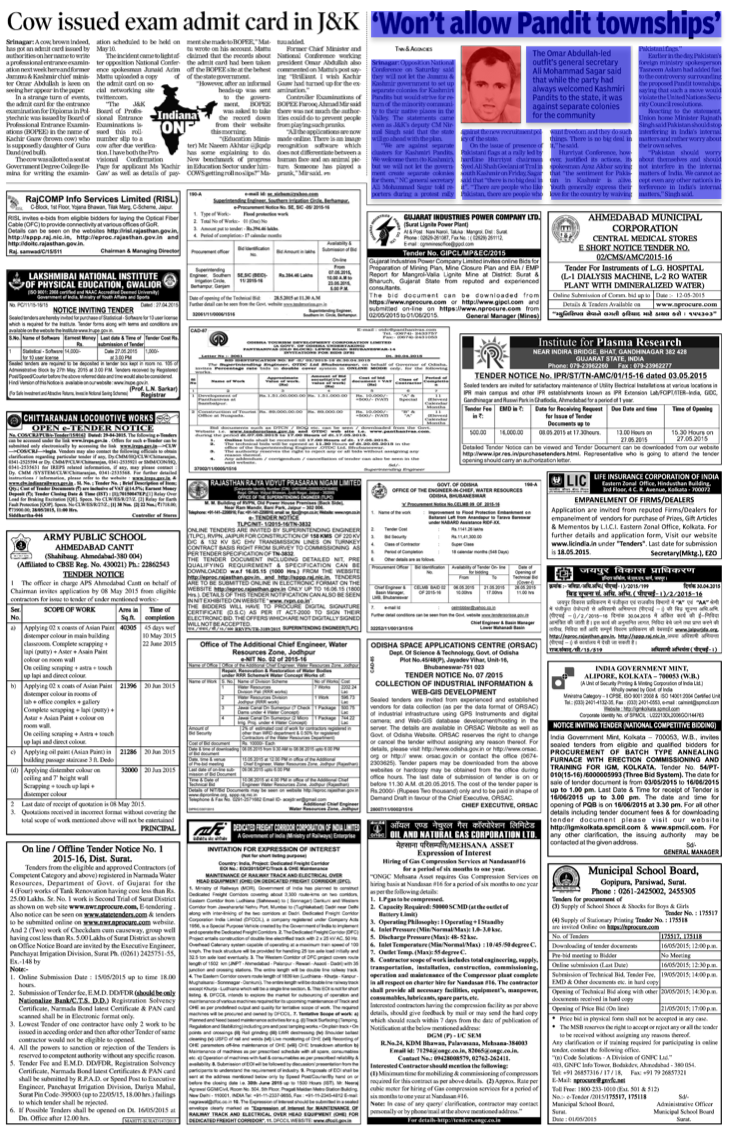}} \hspace{0.5mm}
\caption{Our system is invariant to scale and translation: an
example. The first image is the query image and the second one is
the retrieved image from the database, whose blocks are
of different sizes, aspect ratios and are present at
absolute positions different from those in the query}
\label{fig:Invariance}
\end{figure}

The above systems work on the basis of the layout of the complete
document image, and not specific sub-layouts.
Shin and Doermann \cite{Shin:06} describe a system for sub-layout based document
matching. They measure the query-to-document similarity by
comparing the edges of blocks at
approximately the same location in the query and the candidate
image, after \emph{uniform} scale normalization.
Thus, a solution is
required for the problem of sub-layout-based search where the
query layout can be present at different scales and translations.
Fig. \ref{fig:Invariance} shows an example of query image and the
retrieved image from the database using our method, whose layout
entities are of different size, aspect ratio and the layout
itself is at a position different from the layout in the query image. 
For images in this paper, cyan/blue represents text, pink/red
represent non-text non-background blocks, and grey indicates that
the specific block type (text/non-text) is irrelevant.

Luqman et al. \cite{Luqman:2013} present a graph-based method for document
retrieval, which can perform sub-graph search and spotting. 
The authors start from constructed graphs, and do not consider
segmentation of images and obtaining the subsequent graph structure.
The paper mentions that the overall accuracy depends on the formation of
the graph structure. 
\subsection{Overview and Contributions}
\label{sec:Overview}
In this paper, we present a complete system for layout/sub-layout
based document image retrieval based on the specification of one
or more sub-layout (as a sketch-based query). 
Each database image is processed in an offline process to extract
significant homogeneous regions of layout entities. The processed
image is represented as 2D-graph, where each layout entity
corresponds to graph node and edges are represented by the
relationship between different nodes.
The process starts with the
extraction of layout features in two stages: (a)
Symmetry-maximization-based pre-processing for the cases of
over-segmentation, (b) Generating multiple plausible segmentation
hypotheses from domain specific information. The matching algorithm
uses neighborhood information as a feature in hash-based indexing
for faster retrieval.

The major contributions of this paper are as follows:
\begin{enumerate}
\item The system is script-independent
and works well even for highly unstructured documents such as
newspaper pages.
\item The method works for both complete and partial document matching,
non-uniform scaling and translations.
\item Modeling sketch-based sub-layout queries,
including queries with missing blocks.
The system handles combinations of
multiple sub-layouts, specified as a Boolean query with optional
approximate positional information.
\item A symmetry maximization-based scheme for cases of 
over-segmentation, and using multiple domain-speific segmentation
hypotheses, to increase recall statistics.
\item A hashing-based strategy to prune a possibly large search
space, using neighbourhood information.
\end{enumerate}

The layout of the rest of the paper is as follows. Section 2
describes the preprocessing steps for the extraction of layout
entities and representing a document image in the form of a graph of layout
entities. This section proposes two important methods for handling
the cases of over-segmentation errors and varied range
of user-specified sketch-based queries. Section 3 presents
various types of queries handled by our system and explains
formulation with examples. Section 4 explains in detail our
proposed search and retrieval procedure. This section shows the
use of context information as a feature for hash-based indexing
for pruning the plausibly large search space. Section 5 shows
results of successful retrieval on various types of queries, with
multiple sub-layouts and on documents with script different from
Latin. We conclude the paper in Section 6.
%
%

\section{Pre-processing and Document Representation}
\label{sec:pre_processing_representation}
\begin{figure}[!ht]
\centering
\subfloat[]{\fbox{\includegraphics[width = 3.5cm, height = 4.5cm]{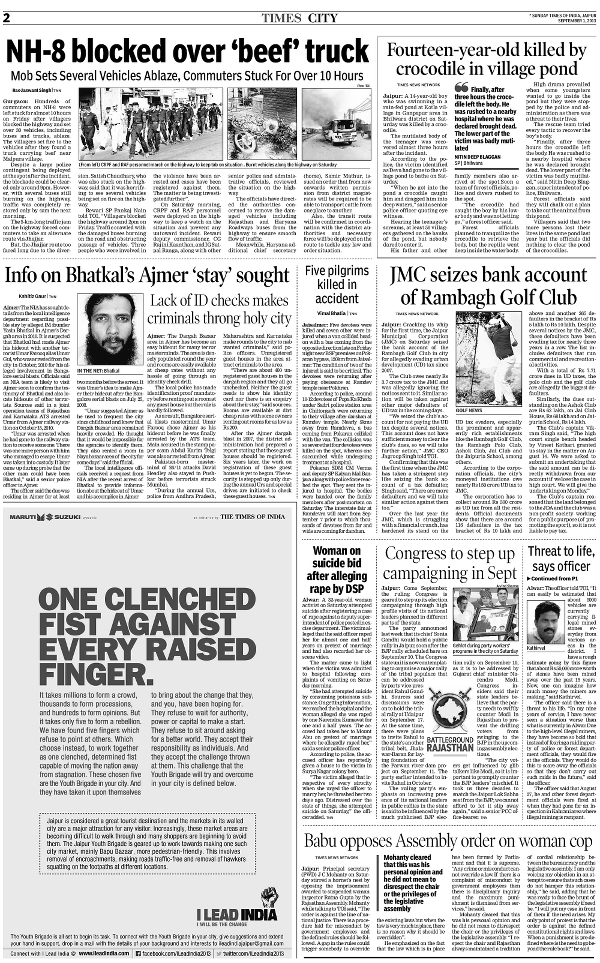}}
\label{fig:Graph_Generation_Original}} \hspace{1mm}
\subfloat[]{\fbox{\includegraphics[width = 3.5cm, height = 4.5cm]{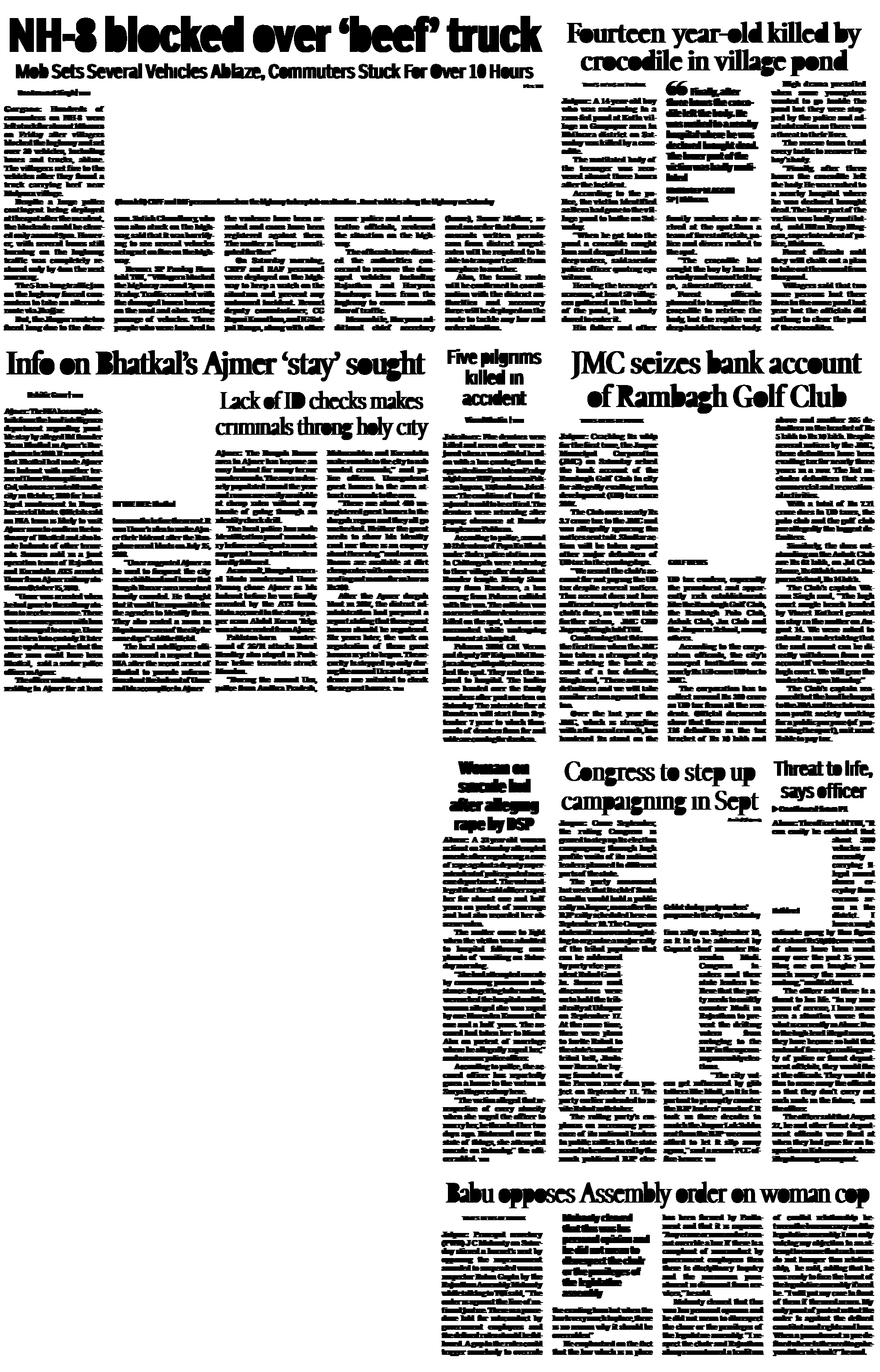}}
\label{fig:Graph_Generation_Seg}} \hspace{0.5mm}\\
\subfloat[]{\fbox{\includegraphics[width = 3.5cm, height = 4.5cm]{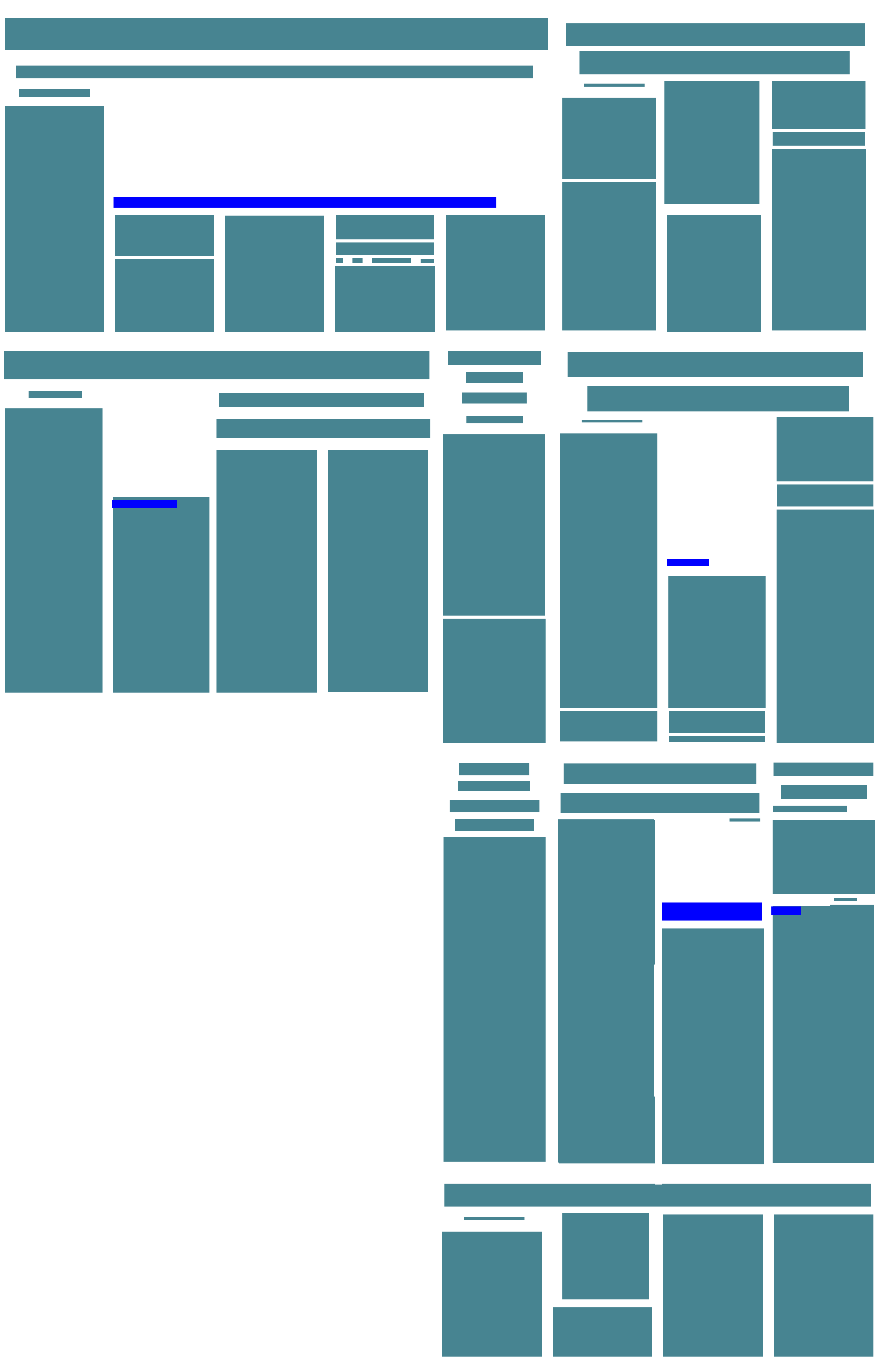}}
\label{fig:Graph_Generation_ARLSA}} \hspace{0.5mm}
\subfloat[]{\fbox{\includegraphics[width = 3.5cm, height = 4.5cm]{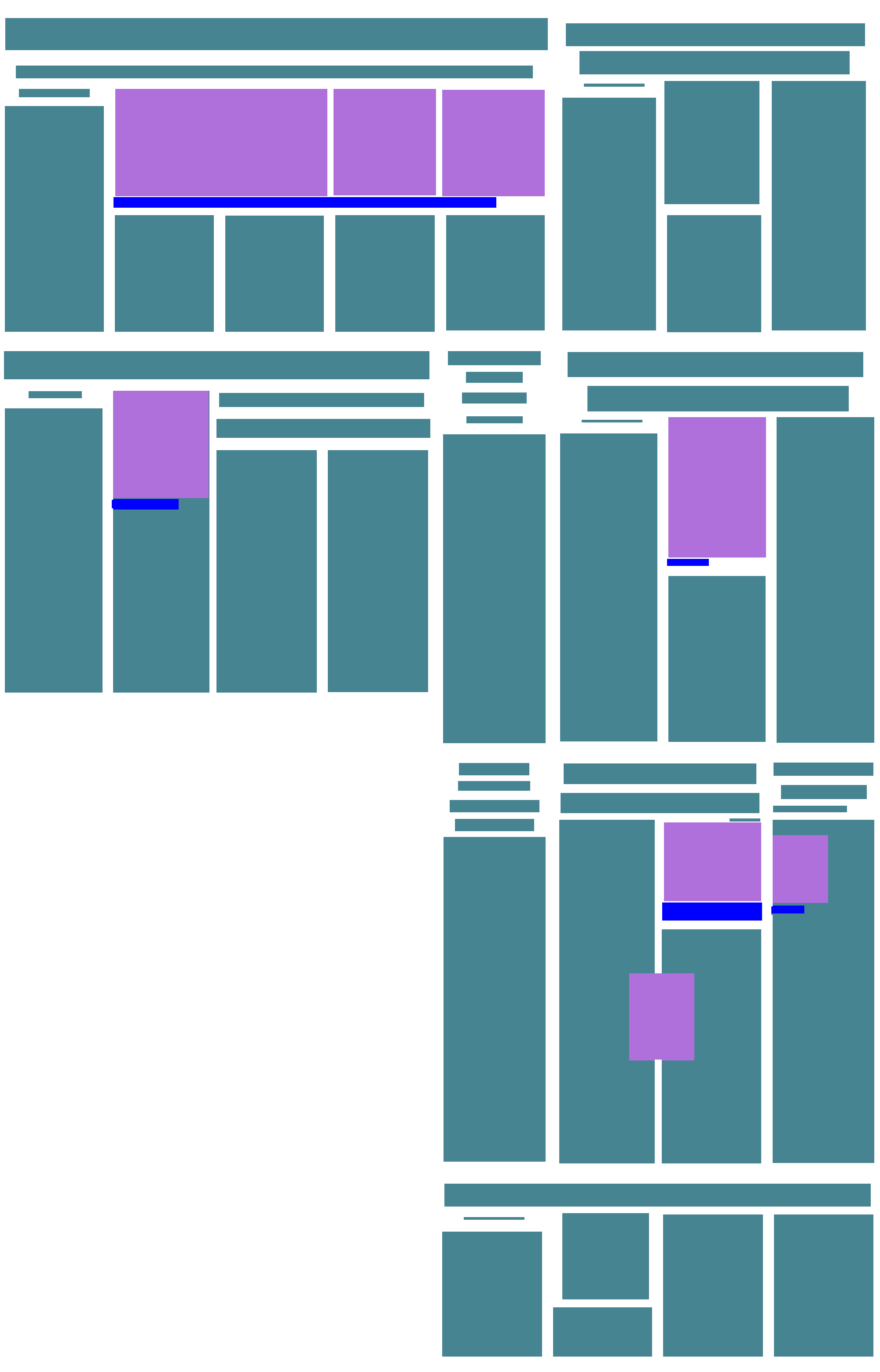}}
\label{fig:Graph_Generation_SymMax}}
\caption{(a) The original image, (b) the output of text-non-text
segmentation, (c) text blocks using ARLSA
(Sec.~\ref{sec:pre_processing_representation}), and
(d) the result of symmetry maximization,
which groups perceptually similar and physically close blocks.
(Sec.~\ref{subsec:symmetry_maximization})}
\label{fig:Graph_Generation_Preprocessing}
\end{figure}

\begin{figure}[h]
\centering
\fbox{\includegraphics[width = 3.5cm, height = 4.5cm]{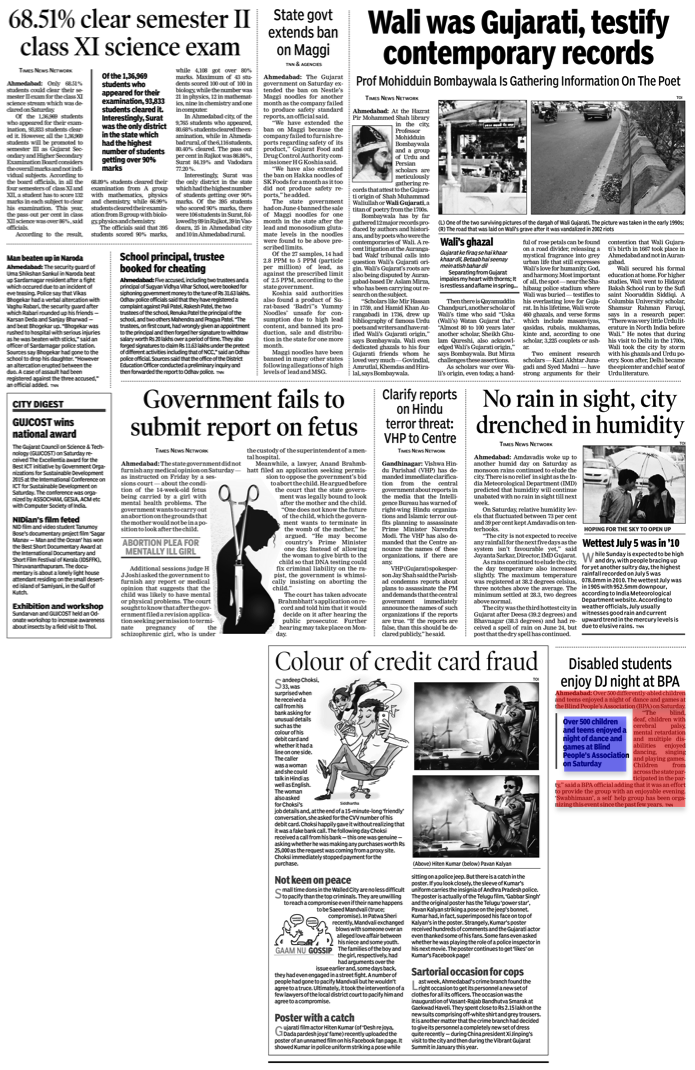}} \hspace{1mm}
\fbox{\includegraphics[width = 3.5cm, height = 4.5cm]{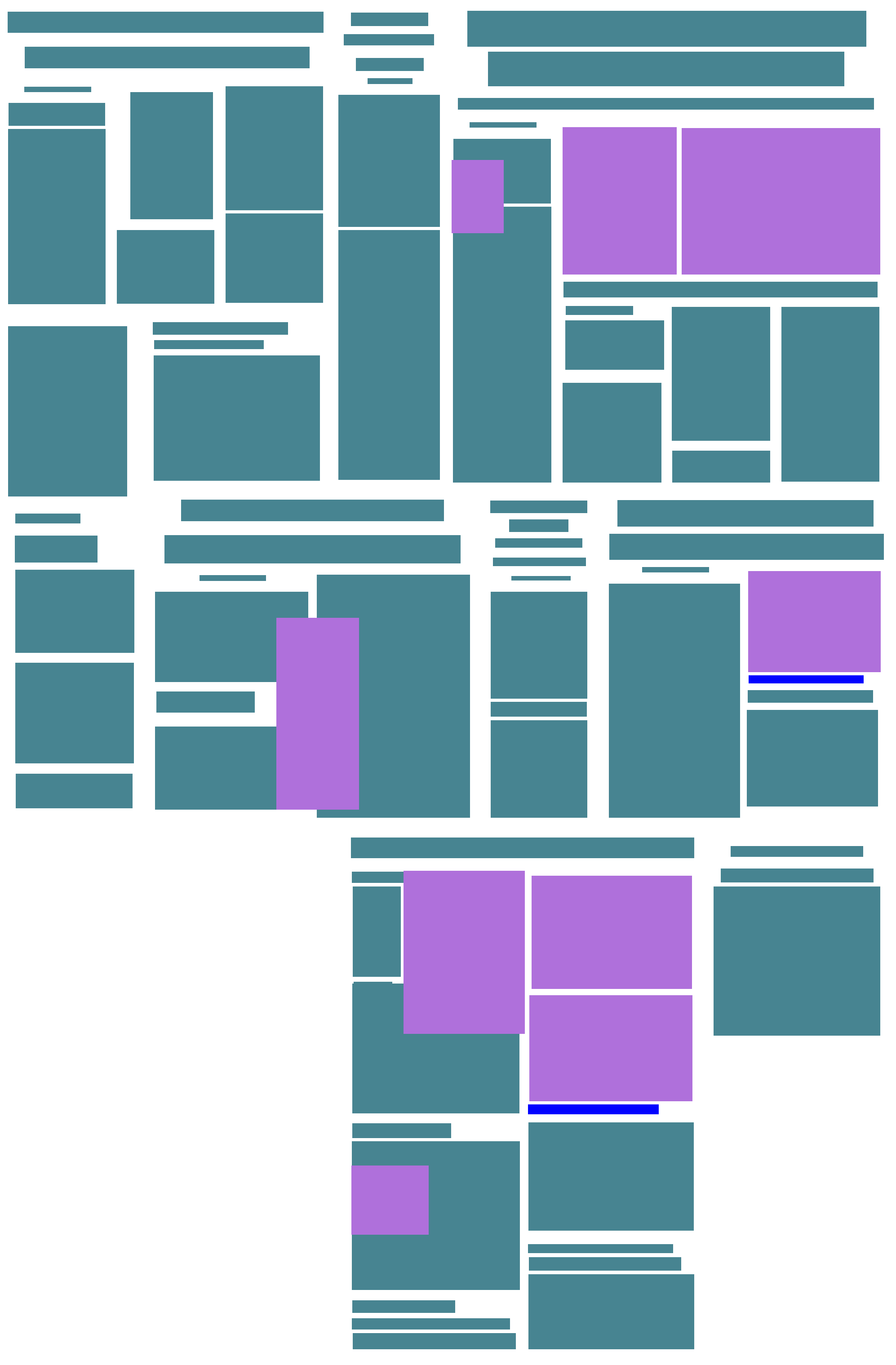}} \hspace{0.5mm}
\caption{ARLSA failure case
(Sec.~\ref{sec:pre_processing_representation}): In case of overlapping text
blocks (shown in red color in the first image), when the minimum
bounding rectangular boxes are formed (shown in second image),
the blocks get merged into one.}
\label{fig:SymmMax_Failure}
\end{figure}

The inputs to the system are scanned document images.
The images are converted into gray scale and are binarized using
Otsu's method \cite{Otsu:1979}. 
Horizontal and vertical lines are identified
using connected component analysis and are removed by replacing
them with background pixels. 
Next, segment the binarized image
into text and non-text regions using multi-level morphological
image processing operations (\cite{Bloomberg:1991}) 
(Fig.~\ref{fig:Graph_Generation_Seg}). 
Homogeneous regions of document blocks are extracted from each image.
Since our dataset contains newspaper images with text of
different font sizes, a simple run-length smoothing algorithm
will not give homogeneous regions. To obtain proper blocks of
text regions, we use Adaptive Run Length Smoothing Algorithm
(hereafter, ARLSA) with a structuring element of size in accordance with the
height of the connected components.  The obtained text blocks are
shown in Fig. \ref{fig:Graph_Generation_ARLSA}.
Out of 5128 newspaper images of our dataset, this step gives us
correct blocks in 93.136\% images. A failure case involves
overlapping text blocks, highlighted in red and blue in 
Fig.~\ref{fig:SymmMax_Failure}.
Creation of minimum bounding rectangles merges these blocks.


The image obtained after ARLSA can be represented as a 2-D graph.
Major limitation of a graph-based representation is that they
are sensitive to noise and over-segmentation errors.
The paper proposes two pre-processing methods suitable for
handling such issues, and going
forward with multiple hypotheses corresponding to
different plausible block graphs. The following sections
explain these in detail.

%

\subsection{Symmetry Maximization}
\label{subsec:symmetry_maximization}
The application of ARLSA gives us the block structure as in
Fig.~\ref{fig:Graph_Generation_ARLSA}.  For layout and
sub-layout-based retrieval, we group paragraphs into a single
block by applying symmetry maximization (\cite{Zhang:2013}), 
inspired by the Gestalt Law of Pragnanz (\cite{Wertheimer:1923})
(which emphasizes the existence of
symmetry and regularity during perceptual grouping).  A symmetry-driven
search is performed in the vertical direction for an optimal
grouping of over-segmented blocks.  Block features used for this
purpose are average character height, alignment (left, right and
center), distance between two blocks, and the presence or absence of
horizontal line between the blocks.  We merge neighboring
blocks, aligned along the top/bottom edges, if the following
conditions are satisfied: (1) they are left, right or
centrally aligned, (2) their average character heights are same,
(3) distance between them is less than average character height
of the image, and (4) no horizontal line is present between them.
Fig.~\ref{fig:Graph_Generation_SymMax}
shows the blocks obtained after this step.

\begin{figure*}[!ht]
\centering
\subfloat[Hypothesis 1]{\fbox{\includegraphics[width = 3.5cm, height = 4.5cm]{./fig/Result_SymMax}}
\label{fig:MultiSeg_H1}} \hspace{0.5mm}
\subfloat[Hypothesis 2]{\fbox{\includegraphics[width = 3.5cm, height = 4.5cm]{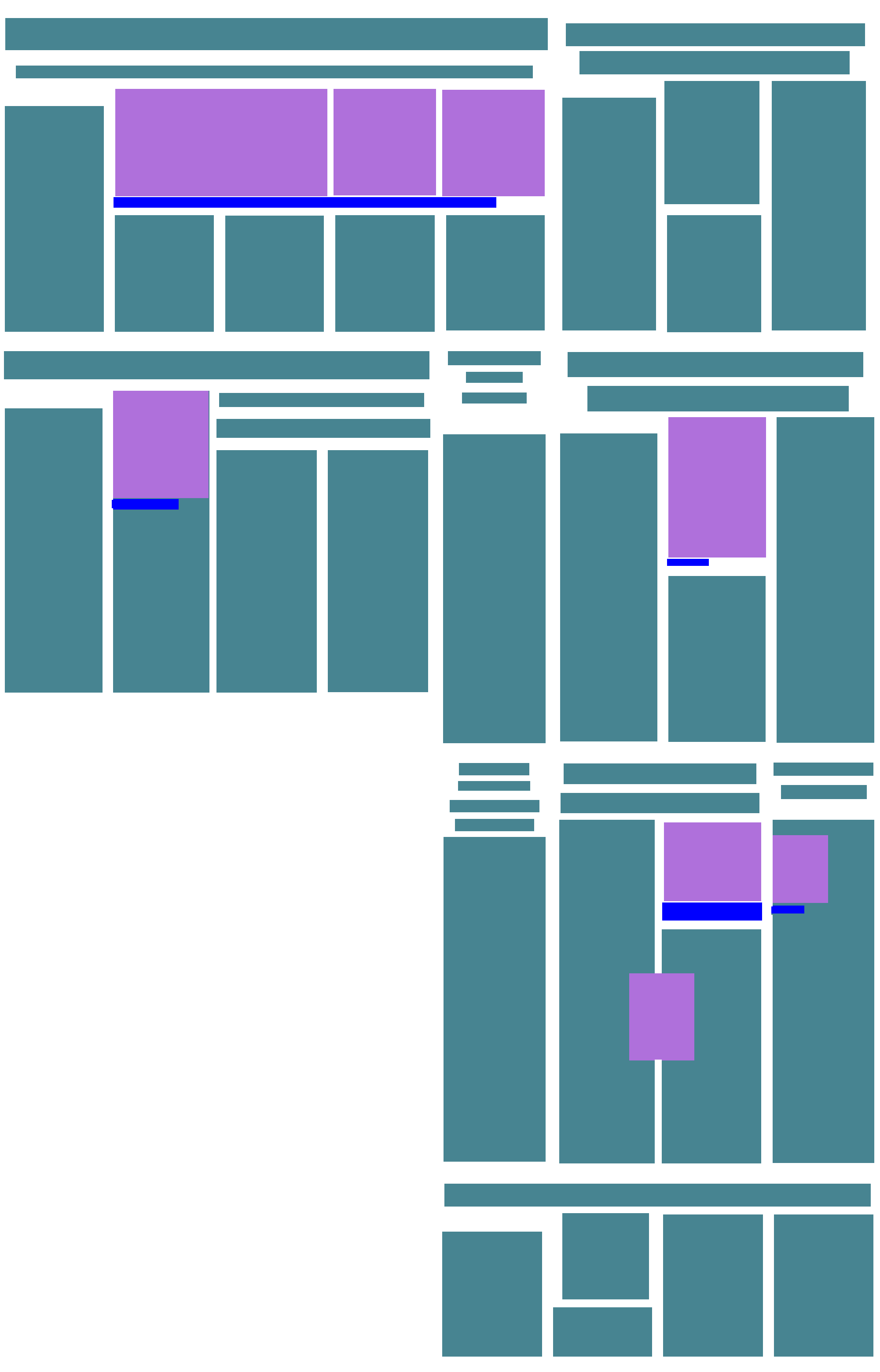}}
\label{fig:MultiSeg_H2}} \hspace{1mm}
\subfloat[Hypothesis 3]{\fbox{\includegraphics[width = 3.5cm, height = 4.5cm]{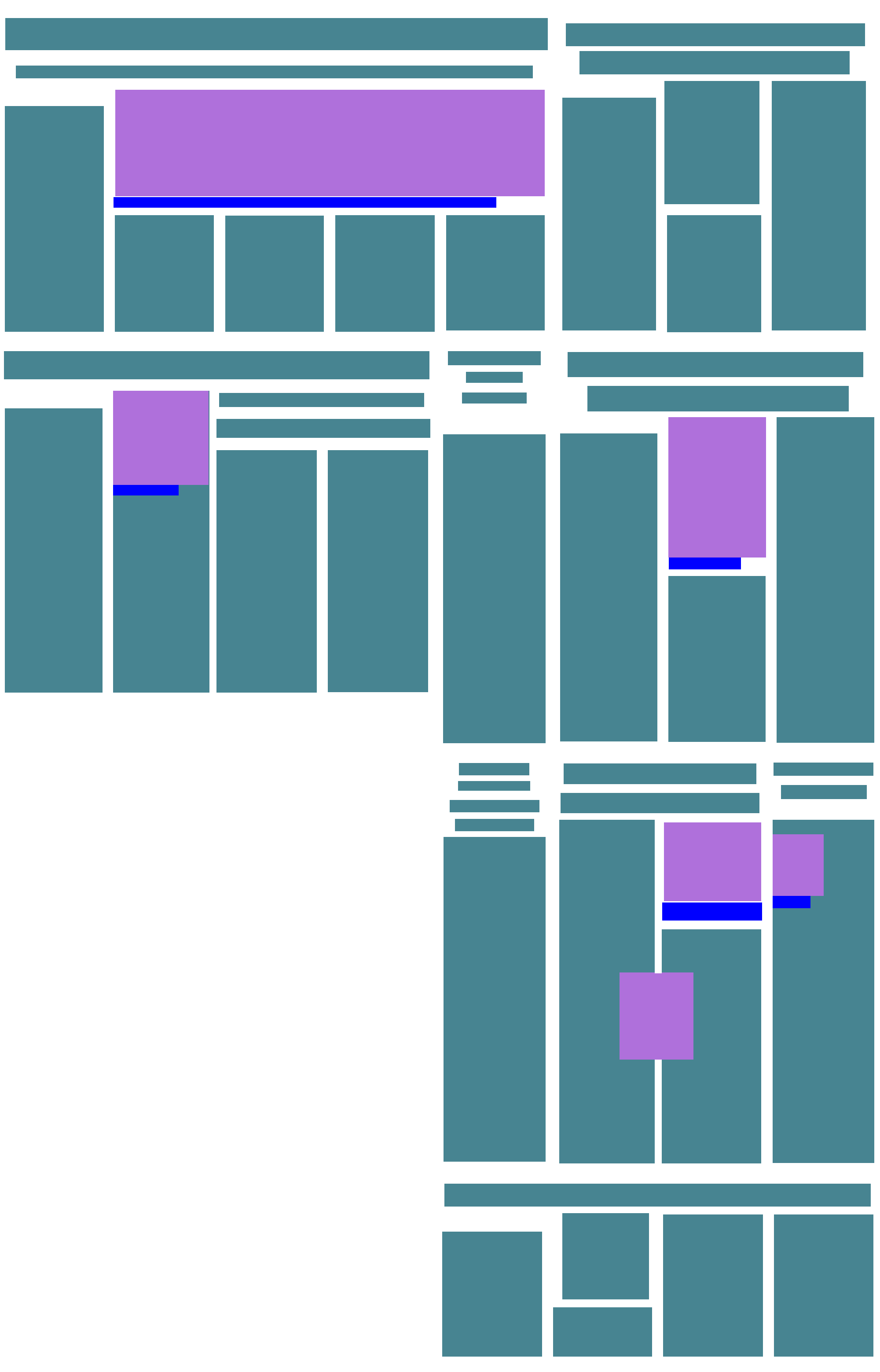}}
\label{fig:MultiSeg_H3}} \hspace{0.5mm}
\subfloat[Hypothesis 4]{\fbox{\includegraphics[width = 3.5cm, height = 4.5cm]{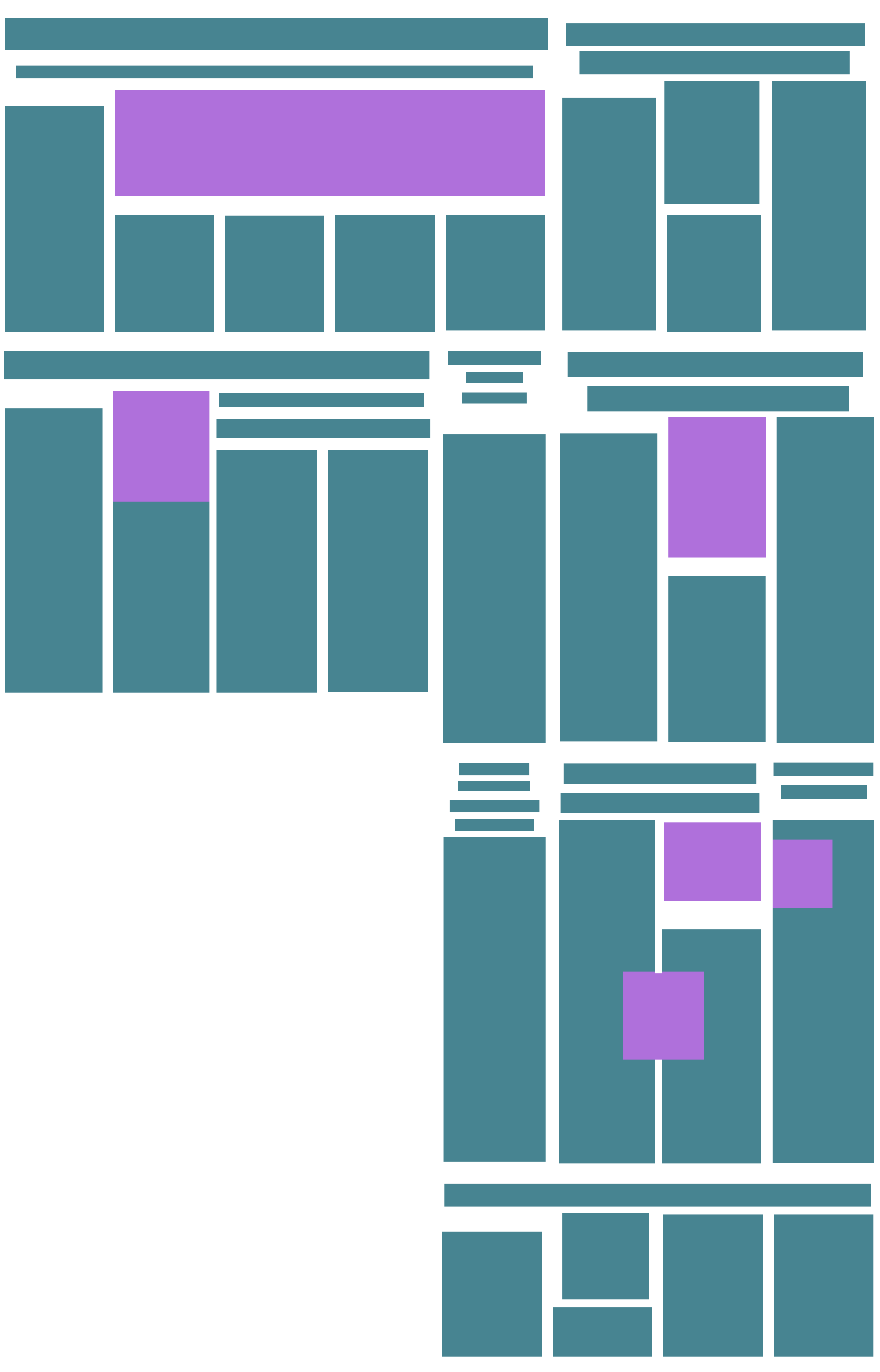}}
\label{fig:MultiSeg_H4}}
\caption{Multiple segmentation hypotheses: 
(a) The ARLSA output (Sec.~\ref{subsec:symmetry_maximization}), 
(b) Hypothesis with small (e.g., noise) or insignificant blocks (author blocks,
for instance) removed,
(c) Hypothesis with grouped adjacent non-text regions, and
(d) Hypothesis with caption blocks removed.}
\label{fig:Graph_Generation_MultiSeg}
\end{figure*}
\subsection{Generating Multiple Segmentation Hypotheses}
\label{subsec:multi_seg_hypotheses}
The graph structure obtained after the application of symmetry
maximization contains blocks of different sizes, some of which
are very small (e.g., author block in newspaper images).  While
drawing blocks for a sketch-based query, a user may not want to
specify small details of the layout. 
\emph{To handle such issues, the system uses domain-specific 
information to get a set of segmentation hypotheses.}
The database stores multiple graphs corresponding to these 
different plausible segmentation hypotheses. 
Fig.~\ref{fig:Graph_Generation_MultiSeg} illustrates this technique.
The first segmentation hypothesis corresponds to the output of
symmetry maximization (Fig.~\ref{fig:MultiSeg_H1}),
explained in the previous section.
Another possible hypothesis considers the removal of small blocks
(which could be noise, or insignificant ones, such as author
blocks). We identify such blocks if their height is
less than or equal to the average character height of the
document image, and which are sandwiched between two text blocks.
Fig.~\ref{fig:MultiSeg_H2} shows such an example.
Another segmentation hypothesis has
merged close-by aligned non-text blocks, as in
Fig.~\ref{fig:MultiSeg_H3} (this example has
three such merged blocks).
Another possible segmentation hypothesis considers
removal of single line caption blocks (on the basis of their
height and position with respect to neighboring blocks),
as in Fig.~\ref{fig:MultiSeg_H4}.
In the database, each block is stored with its attributes 
and context information.
For each block, we store its features such as the height,
width, document ID, block ID, average character height of document,
average character height of block, and spatial location in the
document (top, bottom, left, right, or center). 
Context information consists of the block IDs of the neighboring
blocks towards its four sides i.e., top, bottom, left and right.
This context information is the key feature for hash-based
indexing.  
%


\section{Query Formulation}
\label{sec:Query_Formulation}
\begin{figure*}[!htb]
\centering
\subfloat[Type 1]{\fbox{\includegraphics[width = 1.5cm, height = 2cm]{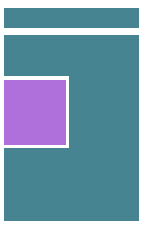}}
\label{fig:query_type1a}} \hspace{0.25mm}
\subfloat[Type 1, again]{\fbox{\includegraphics[width = 2.25cm, height = 2cm]{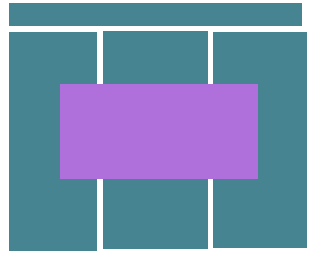}}
\label{fig:query_type1b}} \hspace{0.25mm}
\subfloat[Type 2]{\fbox{\includegraphics[width = 2.25cm, height = 2cm]{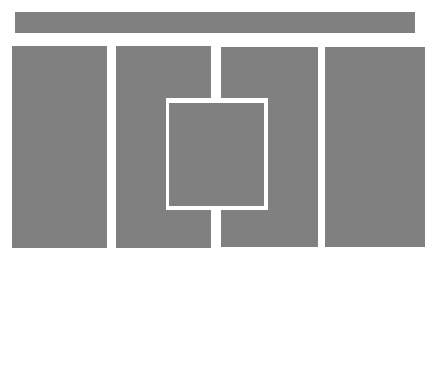}}
\label{fig:query_type2}} \hspace{0.25mm}
\subfloat[Type 3]{\fbox{\includegraphics[width = 2.25cm, height = 2cm]{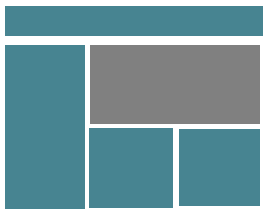}}
\label{fig:query_type3}} \hspace{0.25mm} \\
\subfloat[Type 4]{\fbox{\includegraphics[width = 2.25cm, height = 2cm]{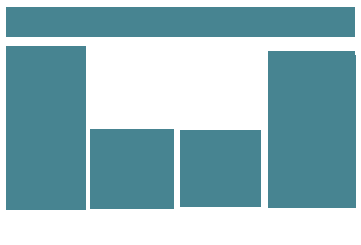}}
\label{fig:query_type4}} \hspace{0.25mm}
\subfloat[Type 5]{\fbox{\includegraphics[width = 2.25cm, height = 2cm]{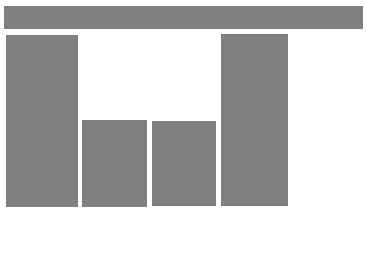}}
\label{fig:query_type5}} \hspace{0.25mm}
\subfloat[Type 6]{\fbox{\includegraphics[width = 1.75cm, height = 2cm]{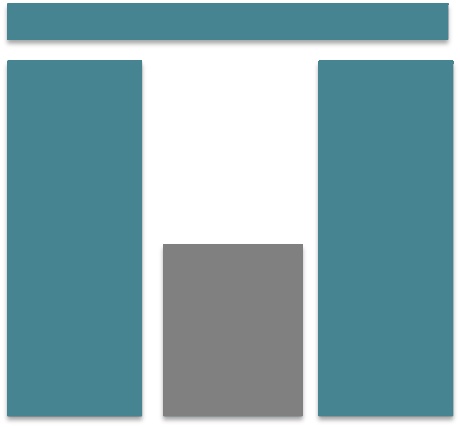}}
\label{fig:query_type6}}
\caption{Various types of query layouts: Blue represents
text, pink represents non-text non-background blocks, and grey
indicates that specific block type (text/non-text) is irrelevant.
(a),(b) All blocks have their type specified, without any missing
blocks, (c) The specific block type (text/non-text) is not relevant
for any block, and there are no missing blocks.
(d) Some blocks need to be retrieved without bothering about
their specific type (text/non-text), and there are no missing 
blocks.
(e) Some blocks missing, with the block type specified
for all blocks.
(f) Missing blocks, and block type (text/non-text) 
is irrelevant for all blocks.
(g) Missing blocks, the type is specified for a few blocks.}
\label{fig:Query_Layouts}
\end{figure*}

\begin{figure*}[!ht]
\centering
\subfloat[]{\fbox{\includegraphics[width = 3cm]{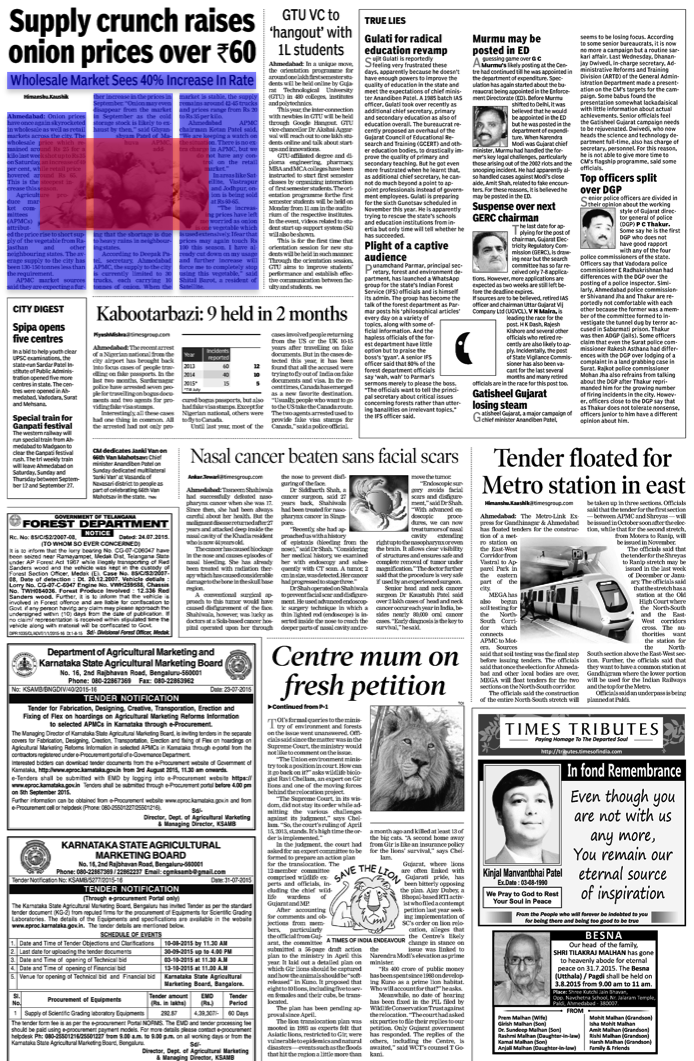}}
\label{fig:R_type1b}} \hspace{0.5mm}
\subfloat[]{\fbox{\includegraphics[width = 3cm]{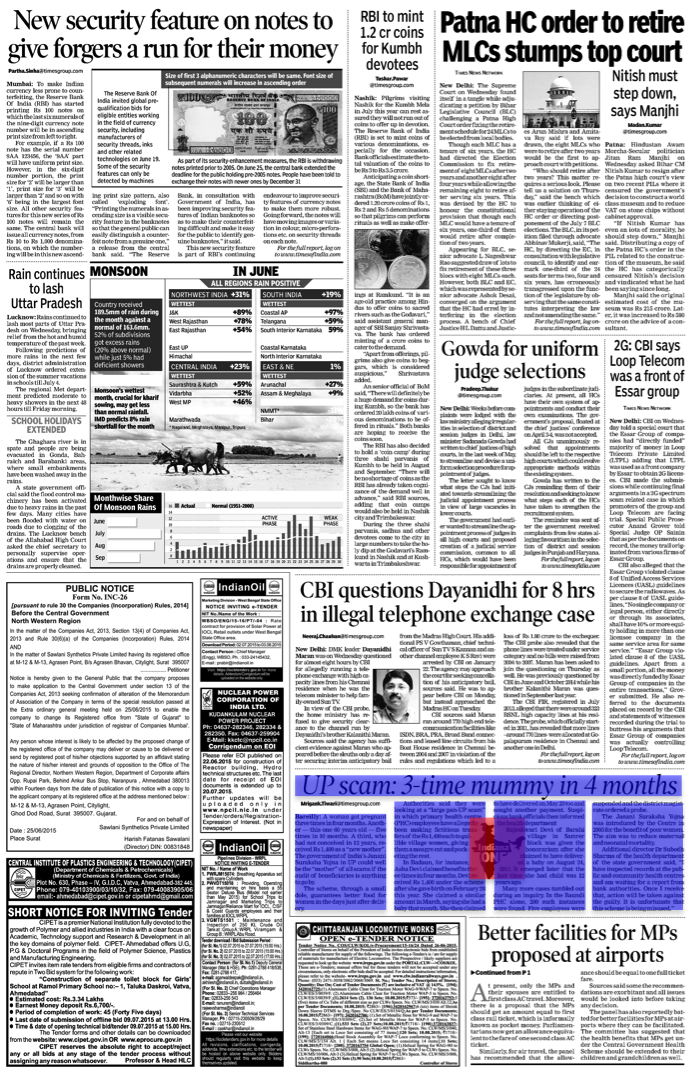}}
\label{fig:R_type2}} \hspace{0.5mm}
\subfloat[]{\fbox{\includegraphics[width = 3cm]{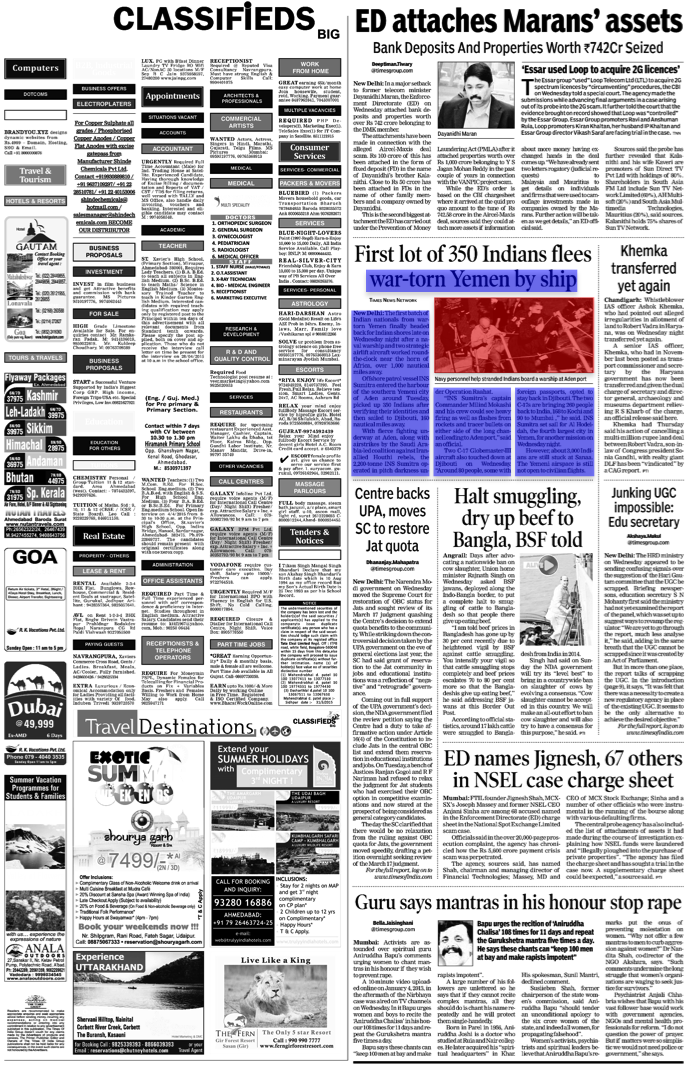}}
\label{fig:R_type3a}} \hspace{0.5mm} \\
\subfloat[]{\fbox{\includegraphics[width = 3cm]{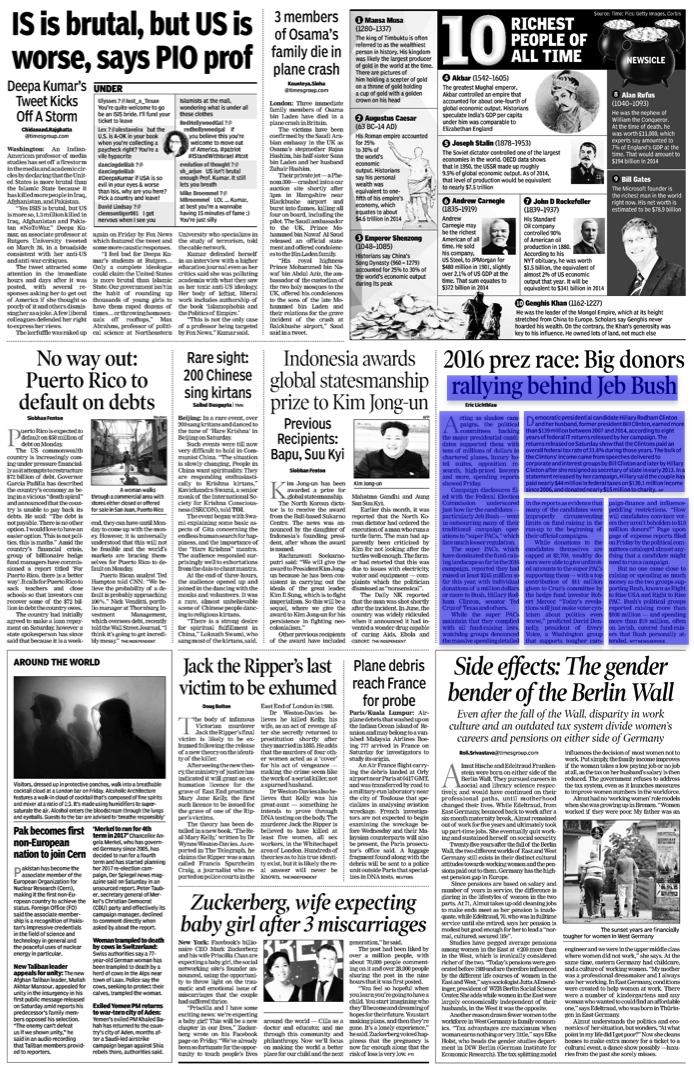}}
\label{fig:R_type3b}} \hspace{0.5mm}
\subfloat[]{\fbox{\includegraphics[width = 3cm]{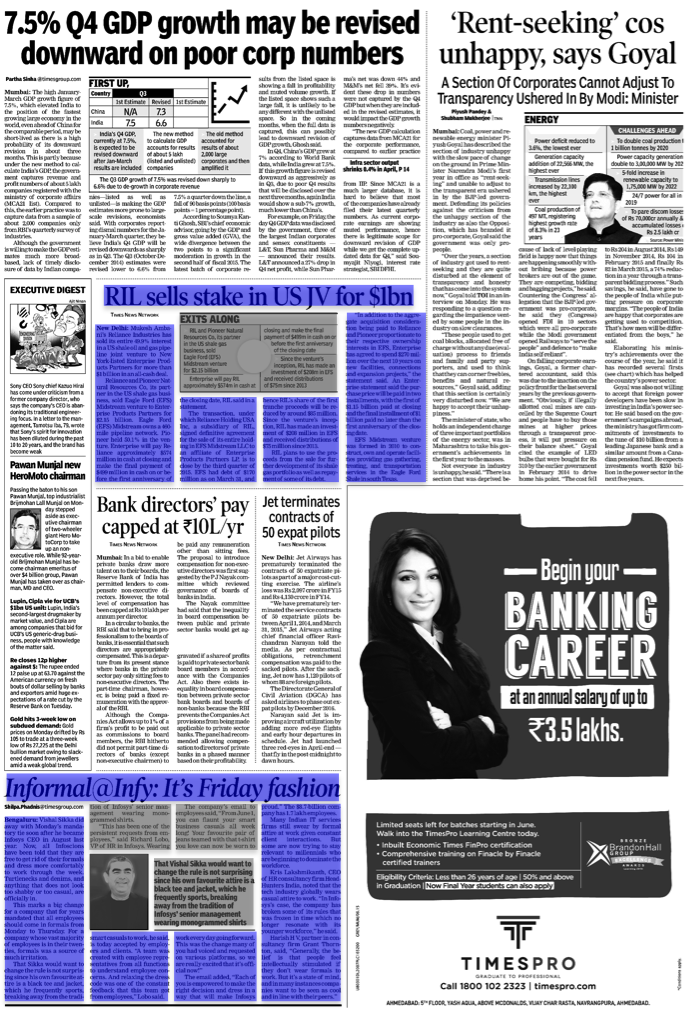}}
\label{fig:R_type4}} \hspace{0.5mm}
\subfloat[]{\fbox{\includegraphics[width = 3cm]{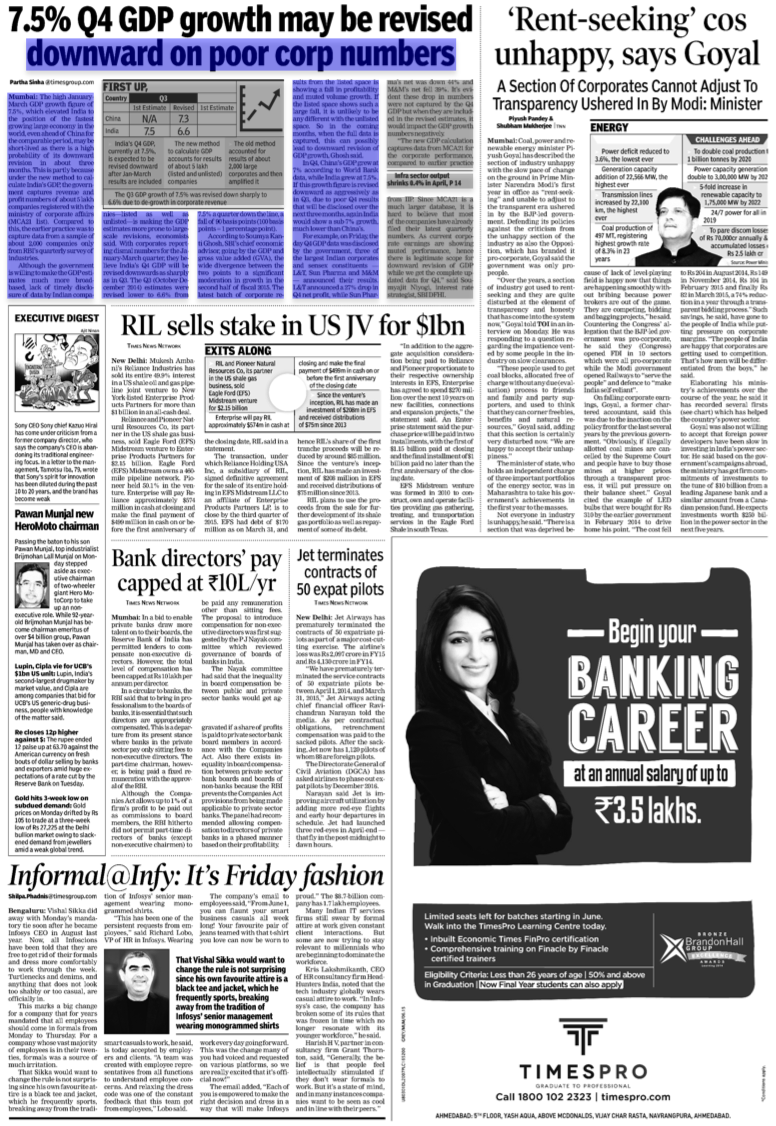}}
\label{fig:R_type5}}
\caption{Retrieved Results: (a) Result for query layout shown in Fig.~\ref{fig:query_type1b}, 
(b) Result for query layout shown in Fig.~\ref{fig:query_type2}. 
For the query layout shown in Fig. \ref{fig:query_type3}, the gray block substituted by non-text block (pink color) in (c) and by text block (blue color) in (d). 
(e) shows the partial matching result for query layout shown in Fig.~\ref{fig:query_type6}.
(f) shows retrieved results for query layout shown in Fig.~\ref{fig:query_type5}: the middle missing block substituted by a non-text one, and the ones to the right, as three text blocks.}
\label{fig:Results}
\end{figure*}

We have conducted a study on 54 people to know their preferences
(what and how) for layout-based searches.
On the basis of their feedback, we have created
the ground-truth, designed queries and adopted a sketch-based method
to specify the desired layout.
Fig. \ref{fig:Query_Layouts} shows the types of queries possible
in our system.
Cyan represents text, and pink, image/non-text/non-background regions. 
Gray indicates cases when the specific block type (text/non-text) 
is not specified by the user.
Our system performs a block arrangement-based match and retrieval.
The user has an option to specify only the desired blocks in a
particular sub-layout.
The remaining space can be left vacant as in
Fig.~\ref{fig:query_type4}, to 
indicate that a \emph{partial match} is intended.
The retrieval system
can then use the vacant space to match with one/more
blocks corresponding to a database document. 
(Fig.~\ref{fig:R_type4}: Sec.~\ref{sec:Results} shows an 
example of retrieval corresponding to the query layout in
Fig.~\ref{fig:query_type4}). 
Queries can be grouped into the following categories:
\begin{enumerate} 
\item Type 1: The type (text/non-text) is specified for all
blocks, without any missing layout entities
(Fig.~\ref{fig:query_type1a},\ref{fig:query_type1b}).
\item Type 2: The type is not specified for any
	block, and there is no missing layout entity 
	(Fig.~\ref{fig:query_type2}).  
\item Type 3: The type is specified for a few blocks, and there
is no missing layout entity (Fig.~\ref{fig:query_type3}).  
\item Type 4: The type is specified for	all blocks, but there are
some missing layout entities (Fig.~\ref{fig:query_type4}).  
\item Type 5: No block has its type specified, and a few layout
entities are missing (Fig.~\ref{fig:query_type5}).  
\item Type 6: The type of a few blocks is specified, and some
layout entities are missing (Fig.~\ref{fig:query_type6}).  
\end{enumerate}


In general, a query layout can be matched anywhere in a document.
\emph{The system also supports retrieval on the basis of a combination
of multiple sub-layouts specified, along with their approximate geometric
locations.} We allow such a combination to be 
conveniently specified using the Boolean operations $AND$, $OR$
and $NOT$.  Consider three sub-layouts $A$, $B$ and $C$, 
of types 1, 2 and 6, respectively.
Suppose the user has specified sub-layout $A$ as in
Fig.~\ref{fig:query_type2}, 
and
Figs.~\ref{fig:query_type1a} and \ref{fig:query_type6} as
sub-layouts $B$ and $C$, respectively. 
The user specifies the required query as:
$(A, bottom)\ AND\ (B)\ AND\ (NOT\ C)$.
In other words,
the user wants to search for documents
which specifically have sub-layouts $A$ and $B$ (with $A$
specifically at the bottom), but not containing sub-layout $C$.
(Sec.~\ref{sec:Results},
Fig.~\ref{fig:Results_Boolean} shows an example of a retrieved
combination of sub-layouts.)
In the next section (Sec.~\ref{sec:Searching_and_Retrieval}), we 
describe the basic search and retrieval strategy for any
particular sub-layout.

\begin{figure*}[!ht]
\centering
\subfloat[]{\fbox{\includegraphics[width = 2.5cm, height = 3.5cm]{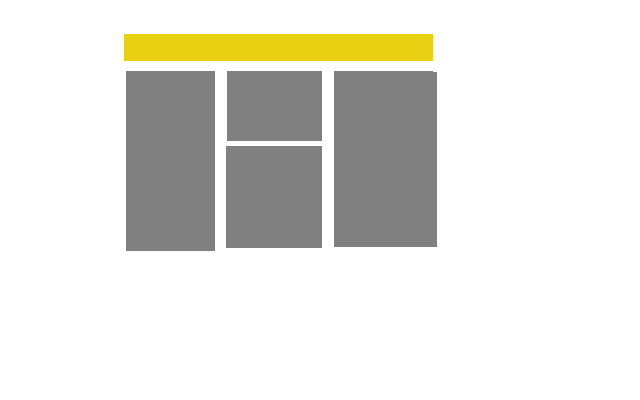}}
\label{fig:Indexing_query}} \hspace{0.5mm}
\subfloat[]{\fbox{\includegraphics[width = 2.5cm, height = 3.5cm]{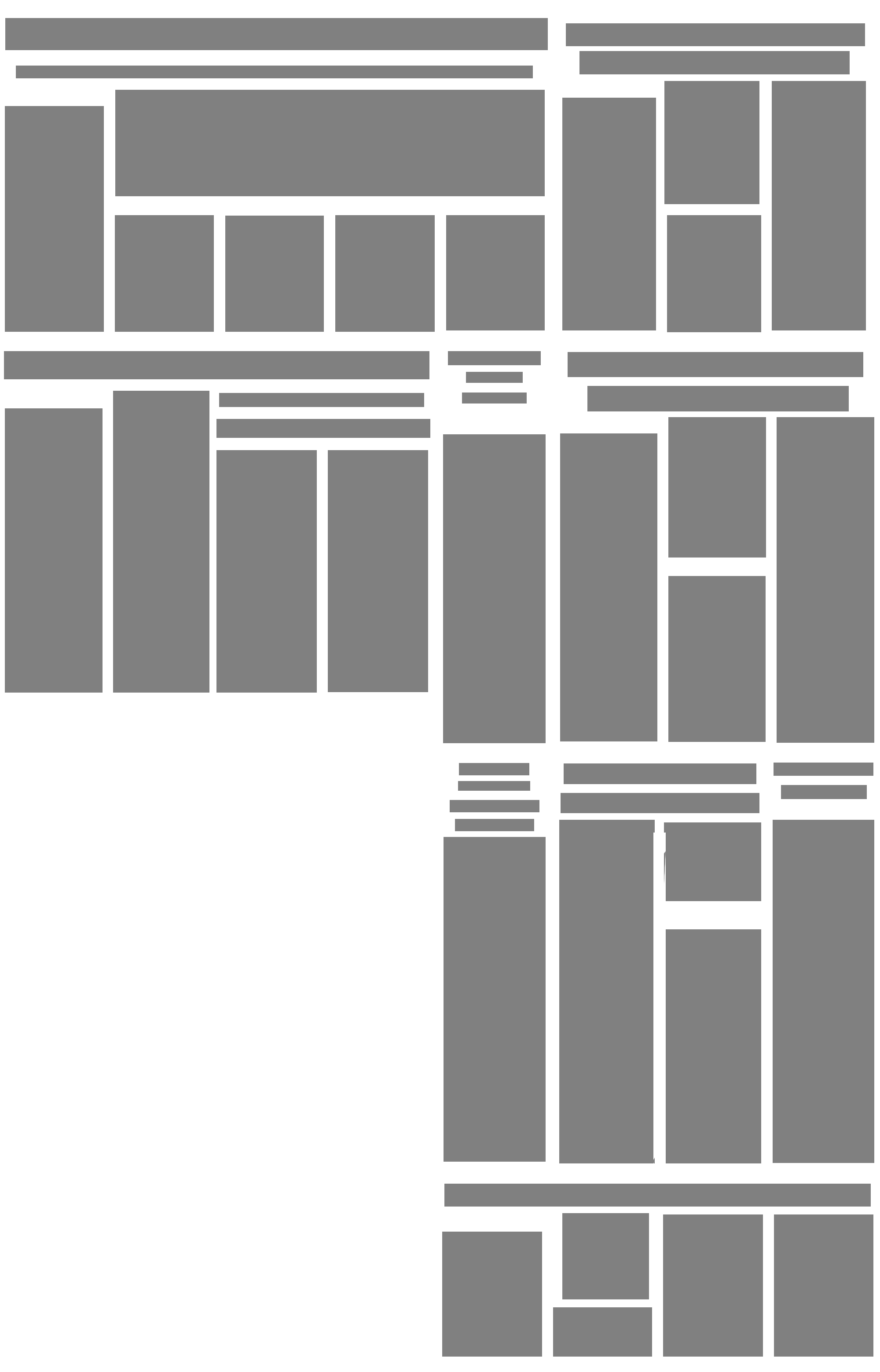}}
\label{fig:Indexing_blockImage}} \hspace{0.5mm}
\subfloat[]{\fbox{\includegraphics[width = 2.5cm, height = 3.5cm]{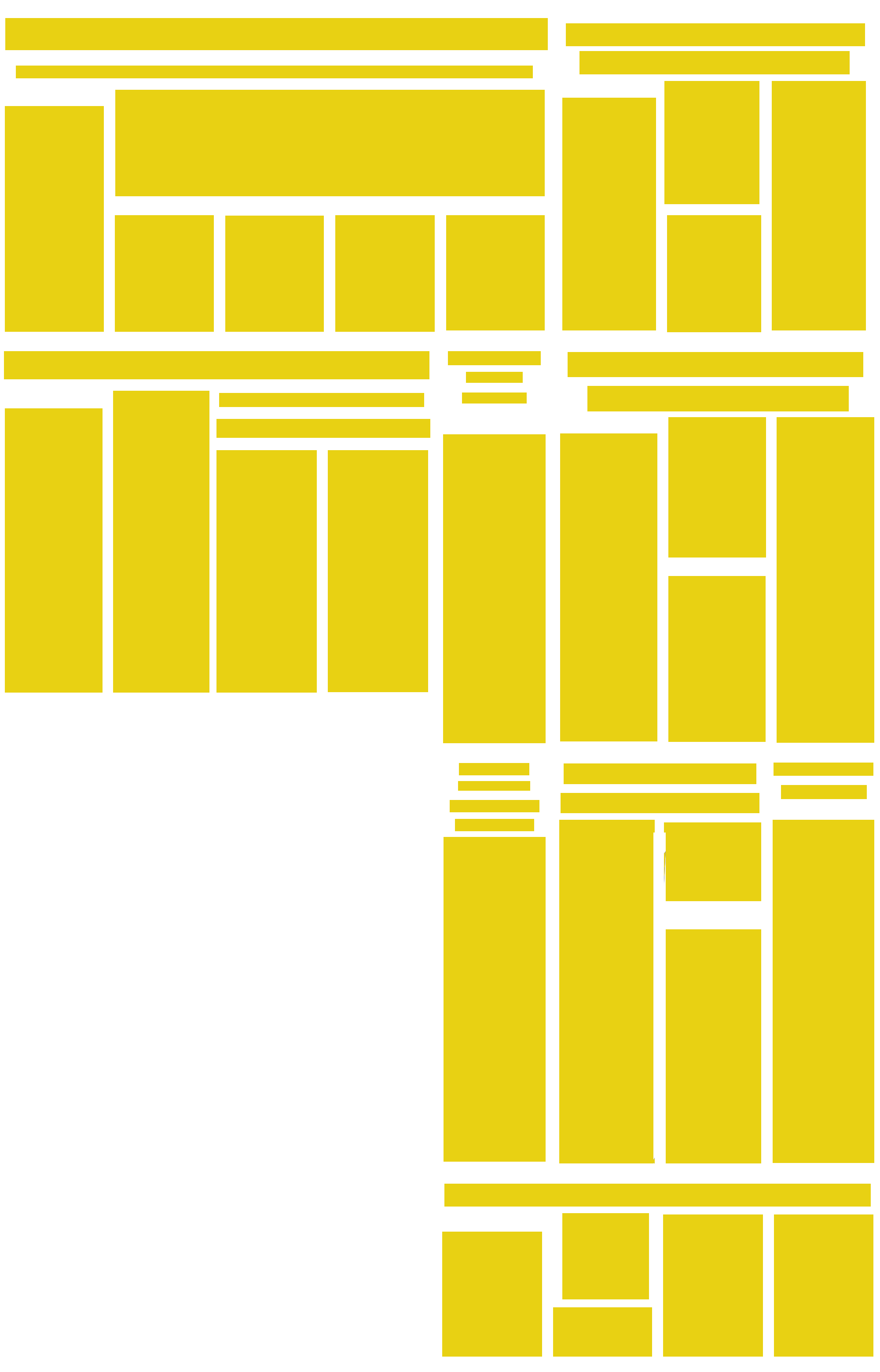}}
\label{fig:Indexing_brute_force}} \hspace{0.5mm}
\subfloat[]{\fbox{\includegraphics[width = 2.5cm, height = 3.5cm]{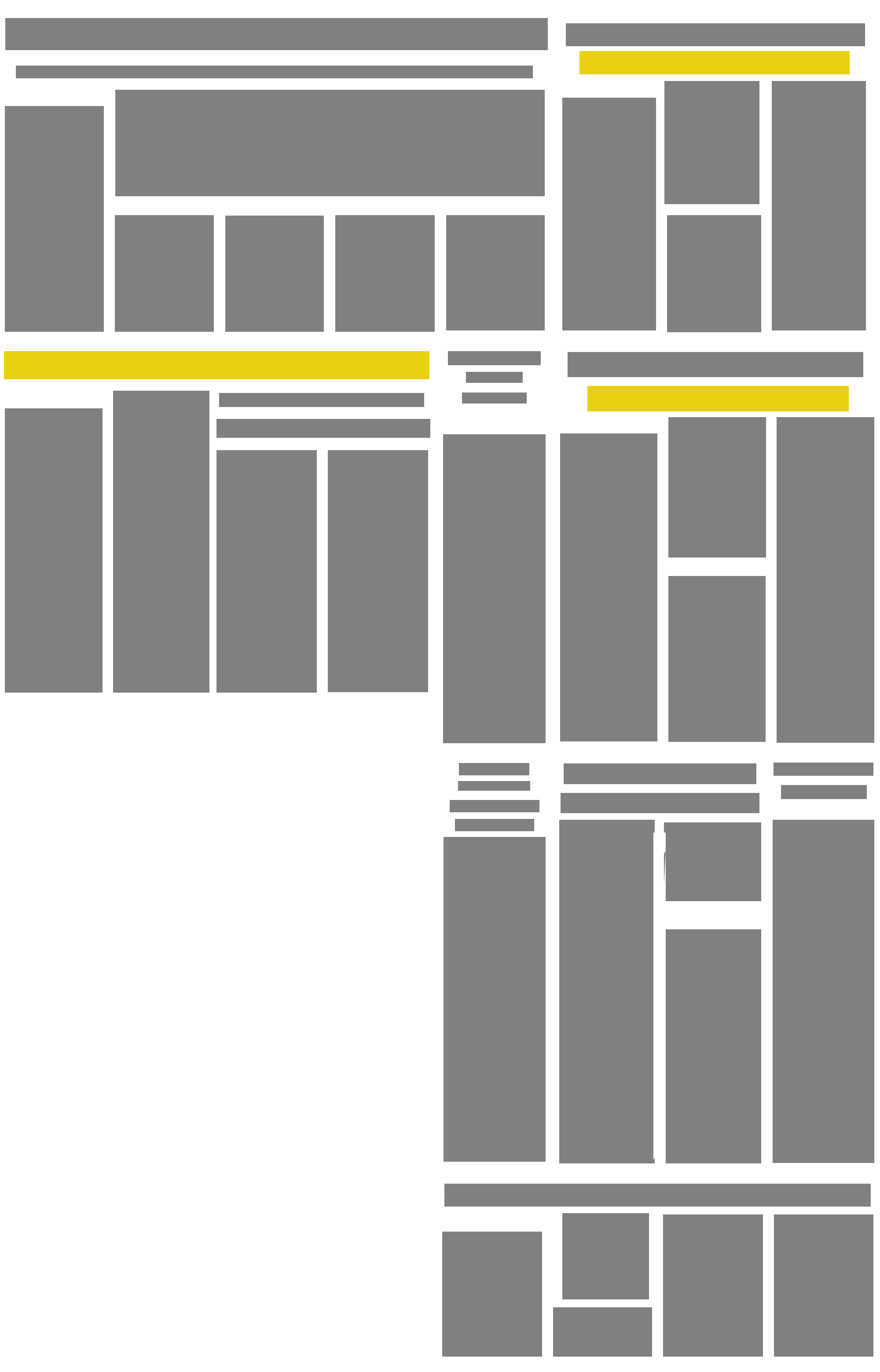}}
\label{fig:Indexing_hash}}
\caption{
(a) The reference block (highlighted in yellow) in the query
image. In this example, without the Hashing-based pruning, 
all blocks (in (b)) get selected as candidate reference blocks
(as shown in (c)). The Hashing-based pruning strategy
(Sec.~\ref{subsec:Indexing}) helps to prune the large search space, as in (d).}
\label{fig:Indexing}
\end{figure*}
\section{The Proposed Search and Retrieval Procedure}
\label{sec:Searching_and_Retrieval}
The proposed graph-based matching
is on the basis of the relative arrangement of blocks,
irrespective of their actual dimensions. 
This imparts relative invariance to factors such as scale and
translation.
The overall steps are as below:
\begin{enumerate}
\item For a database document, 
initialize the candidate reference block set, starting with
all blocks which have a similar neighborhood as the top-left
block of the query (the `reference block', hereafter). 
\emph{An important feature of
our method is to prune this possibly large search space,
using a Hashing-based strategy}. (Sec. \ref{subsec:Indexing} has the details.)
\item Start the matching process with a node from the candidate
reference block set and the query's reference block.
\begin{enumerate}
\item 
Identify if the query sub-layout corresponds to a partial match
(Sec.~\ref{sec:Query_Formulation}).
A partial match considers cases of the dimensions of a vacant
space (due to missing blocks)
being more than a certain fraction of the current block's
dimension (our implementation has 25\%).
Alternately, the dimensions of the vacant space should be
greater than the minimum dimensions of the adjacent blocks.
Fig.~\ref{fig:query_type4} shows a query with a partial match 
at the lower side of the reference block, while
Fig.~\ref{fig:query_type1b} shows a complete match.
\item Compare the neighborhood of the blocks in terms of relative
position and block type (text/non-text). For a partial match,
we insert a dummy block in the vacant space (corresponding to
one or more missing blocks in the query, as in
Fig.~\ref{R_Partial_Query_Dummy}) and match the neighborhood of the 
rest of the blocks.
Fig.~\ref{fig:Results_Ranked} (Sec.~\ref{sec:Results}) shows
final retrieval results involving such missing blocks.
\item Keep traversing the graph till a mismatch is found,
or all blocks of the database document and/or query layout are traversed.
\end{enumerate}
\item If the sub-layout of all non-dummy blocks in the query
matches the sub-layout of blocks corresponding to the candidate
reference block, we declare a match.
\item Go to step 2 and repeat the process with another block from
the candidate reference block set, till all blocks in the set are
processed.
\item The proposed system ranks the retrieved results, in order
of relative average discrepancy in block aspect ratios with
reference to the query sub-layouts, and their relative positions.
Fig.~\ref{fig:Results_Ranked} (Sec.~\ref{sec:Results}) shows an
example of such a ranking.
\end{enumerate}
The proposed method is different from
Coupled Breadth First Search (C-BFS) (\cite{Chikkerur:2006})
in the following ways: (1) by integrating hash-based
indexing, we reduce the time taken in brute force search to a
large extent (explained in Sec.~\ref{subsec:Indexing}), 
(2) Unlike C-BFS, our procedure is capable of handling partial
matches. These situations are practically quite common, 
when a user does not remember the exact layout, or
there are some blocks missing in the query. (3) Our system also ranks the
retrieved results, as explained above.

\subsection{Hashing-based Pruning of the Search Space}
\label{subsec:Indexing}
A brute force-based search of all candidate blocks (as in
C-BFS (\cite{Chikkerur:2006}) for instance),
needs two starting nodes (one from query and one
from database) as reference blocks. 
Since the size of the database is very large, performing one-to-one
matching of blocks is not feasible. In order to narrow down the
portion of database to be explored during the search, we use a
Hashing-based technique. The indexing function predicts the
subset of the database that needs to be searched for each query
image. The context string is extracted for each block, which is
used as the hash key $k$. 
A context string is a fixed-length descriptor that captures the
information of a block (block-type and spatial location in the
document: top, bottom, left or right), and its neighborhood. The
neighborhood information of a block includes the number of
blocks in all four directions (top, bottom, left and right)
and whether they are overlapping with the block.
A hash value is calculated from
it using the equation $ H \: = \; k \; mod \; n $, where
$n$ is the number of bins in the hash table.
The block ID and
database document ID is stored in the
corresponding hash bin. 
Chaining is applied to resolve
collisions that occur when two keys map to the same hash bin. 


Fig.~\ref{fig:Indexing} illustrates how hashing
prunes the search space. Fig.~\ref{fig:Indexing_query} shows a
query layout in which reference node is shown in yellow color.
Fig.~\ref{fig:Indexing_blockImage} is a sample document image
containing 53 blocks.
A brute force technique without hashing will need to consider
each of the 53 blocks as a candidate reference block, as shown in
Fig. \ref{fig:Indexing_brute_force}. 
A good hash-function will put all the blocks
with similar block and context information in the same bucket.
During searching and retrieval, when we match the hash-value
generated using the context string of the query block's reference
node, the number of elements in candidate reference block set gets
reduced to 3 (shown in Fig. \ref{fig:Indexing_hash} in yellow).

\section{Results and Discussion}
\label{sec:Results}
In the absence of a suitable publicly available large dataset, we
created a dataset of newspaper images: a challenging domain, 
since they differ widely in their page layouts,
and subsume more ordered layouts such as in
journals and books, and non-rectangular mixed layouts in some magazines.
ARLSA generates correct blocks on 4776 of 5128 images.
(Details in Sec.~\ref{sec:pre_processing_representation}, including
a failure case example.)
We create a graph
database for these 4776 images, using symmetry maximization
(Sec.~\ref{subsec:symmetry_maximization})
and multiple segmentation hypotheses
(Sec.~\ref{subsec:multi_seg_hypotheses}).
\subsection{Handling different types of Queries}
\label{subsec:different_queries}
\begin{figure}[!ht]
\centering
\fbox{\includegraphics[width = 2.5cm]{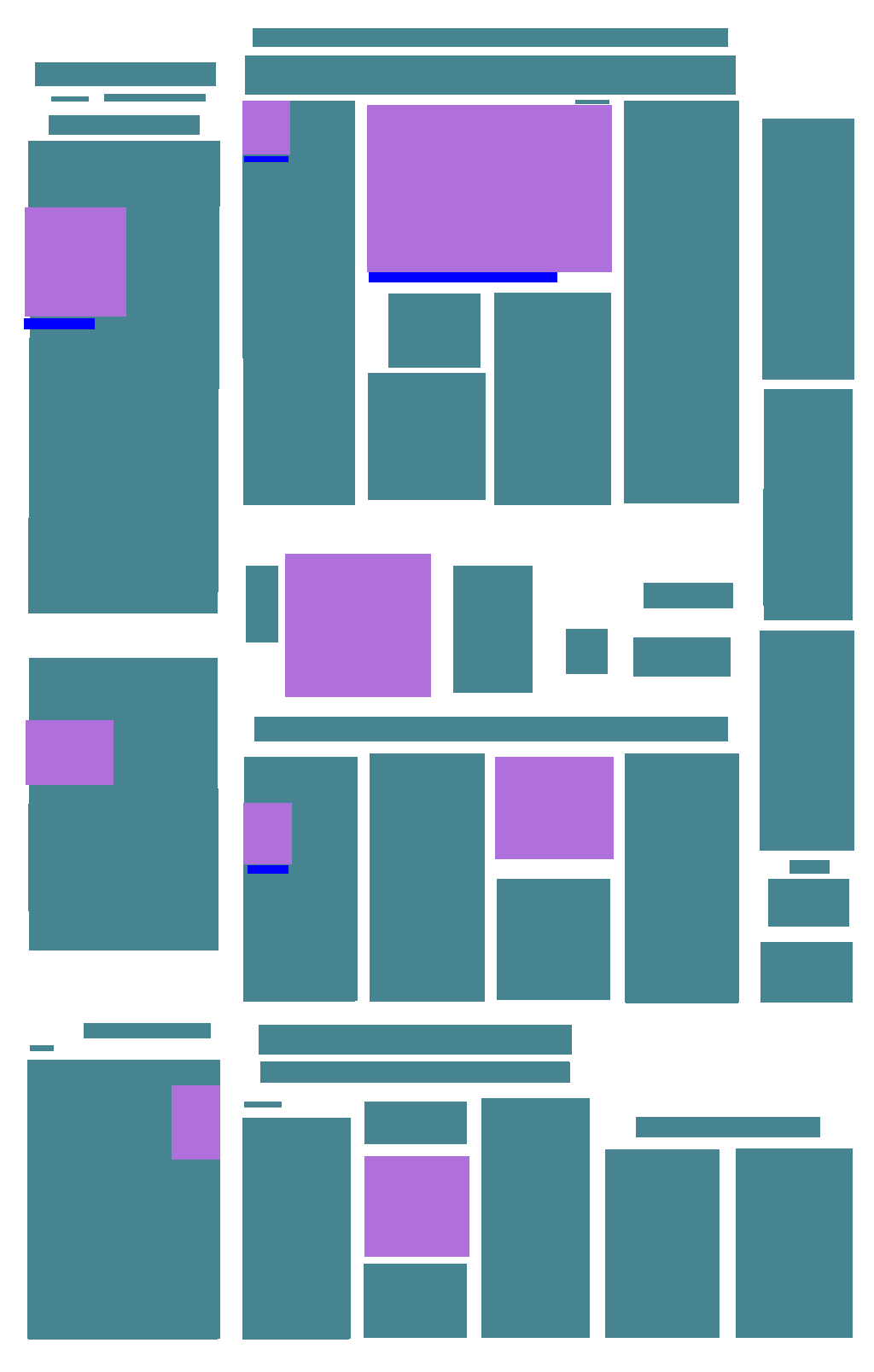}} \hspace{1mm}
\fbox{\includegraphics[width = 2.5cm]{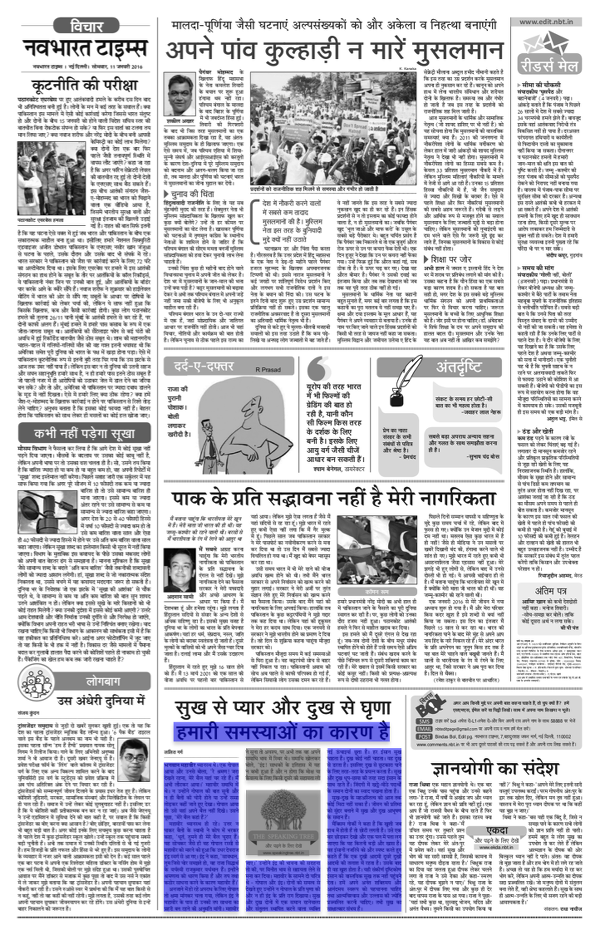}}
\caption{The system is language and script-independent since it
considers the relative arrangement of blocks. For a
Hindi newspaper in Devanagari script, the images above show
the block structure
(Sec.~\ref{sec:pre_processing_representation}), and results of
successful retrieval (Query Type 6, Fig.~\ref{fig:query_type6}).}
\label{fig:results_hindi}
\end{figure}

\begin{figure}[!ht]
\centering
\fbox{\includegraphics[width = 1.5in]{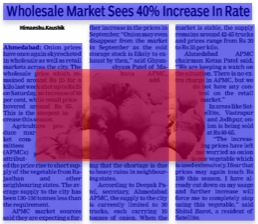}
\label{R_type1b_zoom}} 
\fbox{\includegraphics[width = 1.5in]{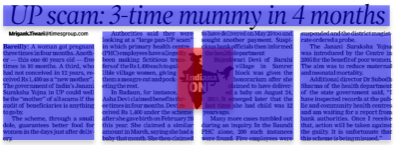}
\label{R_type2_zoom}}
\caption{Handling irregular layouts: the two examples above
correspond to lines of text interspersed with images of irregular
dimensions.
The system considers the closest rectangular bounding boxes and
is able to perform correct retrieval: these correspond to
Figs.~\ref{fig:R_type1b} and \ref{fig:R_type2}.}
\label{fig:Results_Complex}
\end{figure}

Fig.~\ref{fig:Results} summarizes  
experimental results with various types of query layouts
(Fig.~\ref{fig:Query_Layouts}). 
(We describe a case of retrieval with the sub-layout in
Fig.~\ref{fig:query_type1a} specifically in the context of
simultaneous retrieval of multiple sub-layouts, later in this section).
A retrieval example for the sub-layout Type 1b 
(Fig.~\ref{fig:query_type1b}): block types
completely specified, no blocks) is in
Fig.~\ref{fig:R_type1b}. For sub-layout Type 2 (Fig.~\ref{fig:query_type2}: no
block type specified, no missing blocks), Fig.~\ref{fig:R_type2}
shows an example of successful retrieval.
In the query layout of Type 3 (Fig.~\ref{fig:query_type3}), the
gray block indicates that the block type (text/non-text) has not
been specified.
The retrieval result in Fig.~\ref{fig:R_type3a} shows the gray
block substituted by a non-text block (highlighted in pink),
In Fig~\ref{fig:R_type3b}, it is substituted by a text block
(highlighted in blue).
Query Type 4 (Fig.~\ref{fig:query_type4}) has all block types
specified but missing blocks. Fig.~\ref{fig:R_type4} 
shows a retrieval result with the same sub-layout found in two
places in the same document: the top one has one missing non-text
block, while the bottom one has four missing blocks (3 text and
one non-text). All missing blocks in
Fig.~\ref{fig:R_type4} are shown in gray.
In Query Type 5 (Fig.~\ref{fig:query_type5}), 
the type (text/non-text) is not specified for
any block, and there are one or more missing blocks,
which may be spatially located anywhere. Fig.~\ref{fig:R_type5}
shows a retrieved result, which has the middle missing block as a
non-text one, and the ones to the right, as three text blocks.
Query Type 6 (Fig.~\ref{fig:query_type6}) considers cases with the
block type specified for a few blocks, and missing blocks.
Fig.~\ref{fig:results_hindi} shows an example of successful
retrieval interestingly, for a newspaper in a different language
and script (Hindi, Devanagari). \emph{The system is
independent of the specific language and the script, since the
retrieval is on the basis of the relative arrangement of blocks.}
Fig.~\ref{fig:Results_Complex} shows results of successful
retrieval of layouts with lines of text interspersed with parts of images.
(These correspond to Fig.~\ref{fig:R_type1b} and Fig.~\ref{fig:R_type2}.
The system considers the minimum rectangular bounding boxes for
the irregular images. As mentioned in
Sec.~\ref{sec:pre_processing_representation}
(Fig.~\ref{fig:SymmMax_Failure}), the system fails as the
pre-processing module is not able to handle overlapping text
blocks. As shown in Fig.~\ref{fig:Results_Complex}, the system
can handle overlapping blocks of different types (text and
non-text).
\subsection{Handling Combinations of Multiple Sub-layouts}
\label{subsec:R_Boolean}
\begin{figure*}[!ht]
\centering
\fbox{\includegraphics[width = 2.5cm]{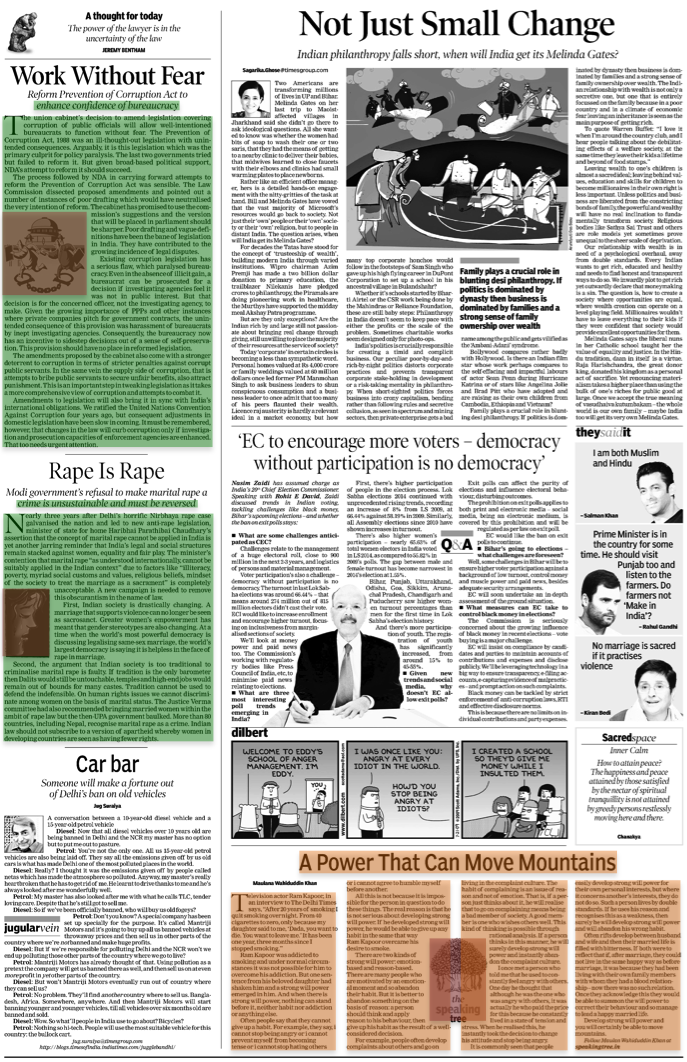}
\label{R_Boolean1}}
\fbox{\includegraphics[width = 2.5cm]{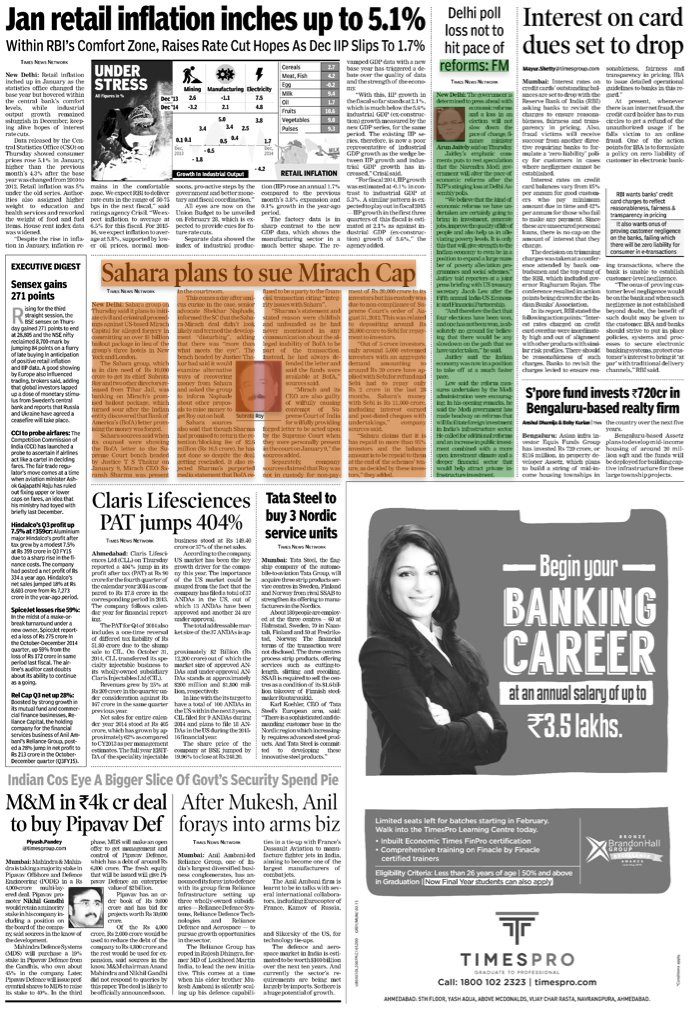}
\label{R_Boolean2}} \hspace{0.25mm}
\fbox{\includegraphics[width = 2.5cm]{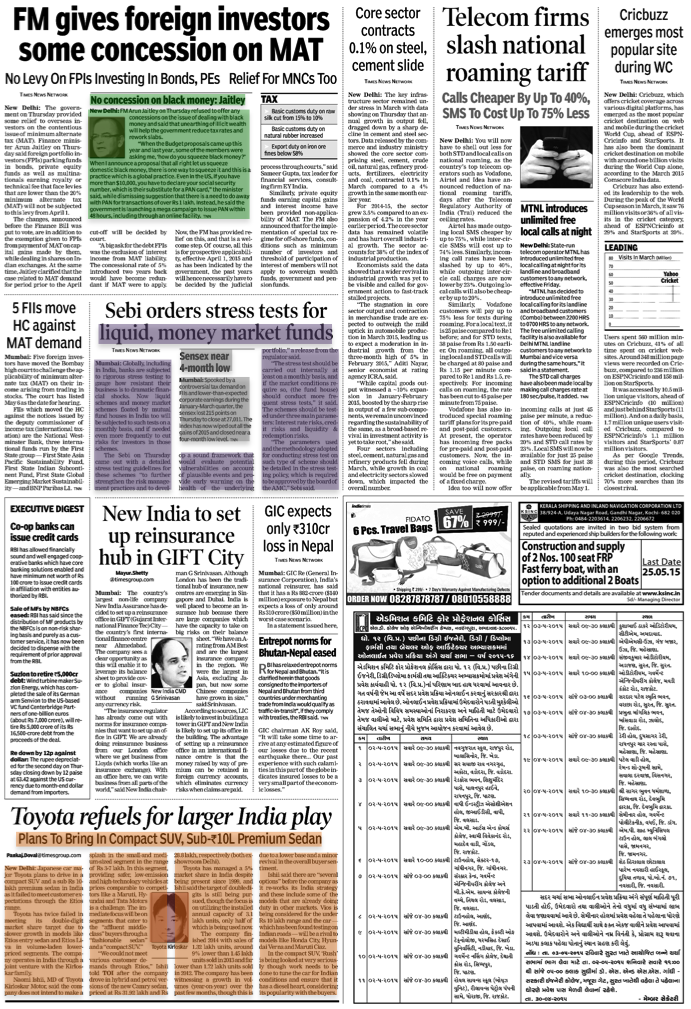}
\label{R_Boolean3}}
 \hspace{0.25mm}
\caption{Retrieval results for a combination of multiple
sub-layouts in different spatial locations
(Sec.~\ref{sec:Query_Formulation}).
Query $(A, bottom)\ AND\ (B)\ AND\ (NOT\ C)$
(Sec.~\ref{subsec:R_Boolean}) results in left-most result alone,
and not the other two. The middle one does
not satisfy the spatial location requirements for sub-layout $A$,
whereas the rightmost one also contains layout $C$.}
\label{fig:Results_Boolean}
\end{figure*}

As mentioned in Sec.~\ref{sec:Query_Formulation}, the system
supports retrieval of a combination of multiple sub-layouts.
Fig.~\ref{fig:Results_Boolean} shows the retrieval results for
the query $(A, bottom)\ AND\ (B)\ AND\ (NOT\ C)$. 
The sub-layouts A, B, and C correspond to query layouts in
Fig.~\ref{fig:query_type1b}, \ref{fig:query_type1a}, and
\ref{fig:query_type6}, respectively. 
The left-most document in Fig.~\ref{fig:Results_Boolean} is the
correctly retrieved document which satisfies all 4 constraints. 
The middle one is not retrieved since it violates the spatial constraint for
sub-layout $A$. The right-most is not retrieved
since the presence of sub-layout $C$ violates the query
specifications.
\subsection{Ranking of Results, and Retrieval Statistics}
\label{subsec:ranking_statistics}
\begin{figure*}[!htb]
\centering
\begin{minipage}[b][3cm][c]{2.5cm}
\subfloat[]{\fbox{\includegraphics[width = 2cm, height = 1.4cm]{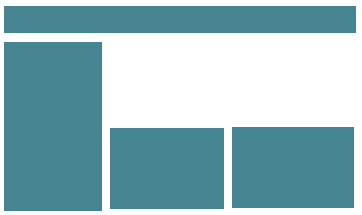}}
\label{R_Partial_Query}} \vspace{.25mm}
\subfloat[]{\fbox{\includegraphics[width = 2cm, height = 1.4cm]{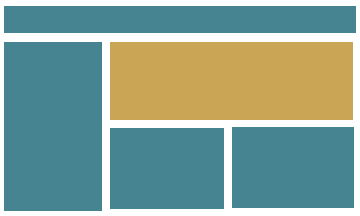}}
\label{R_Partial_Query_Dummy}}
\end{minipage}
\subfloat[]{\fbox{\includegraphics[width = 2.5cm, height = 3.5cm]{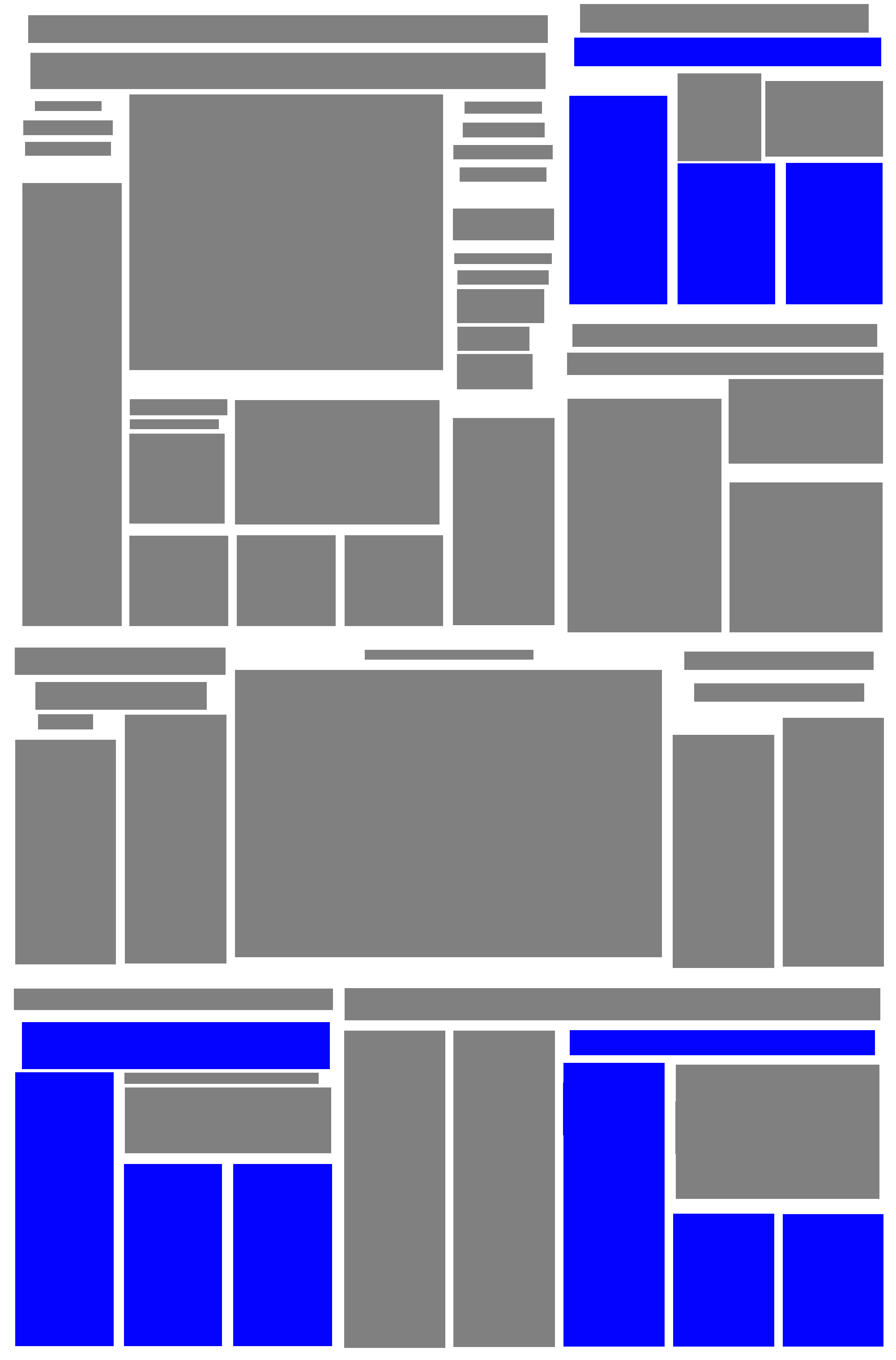}}
\label{R_partial1}} \hspace{0.5mm}
\subfloat[]{\fbox{\includegraphics[width = 2.5cm, height = 3.5cm]{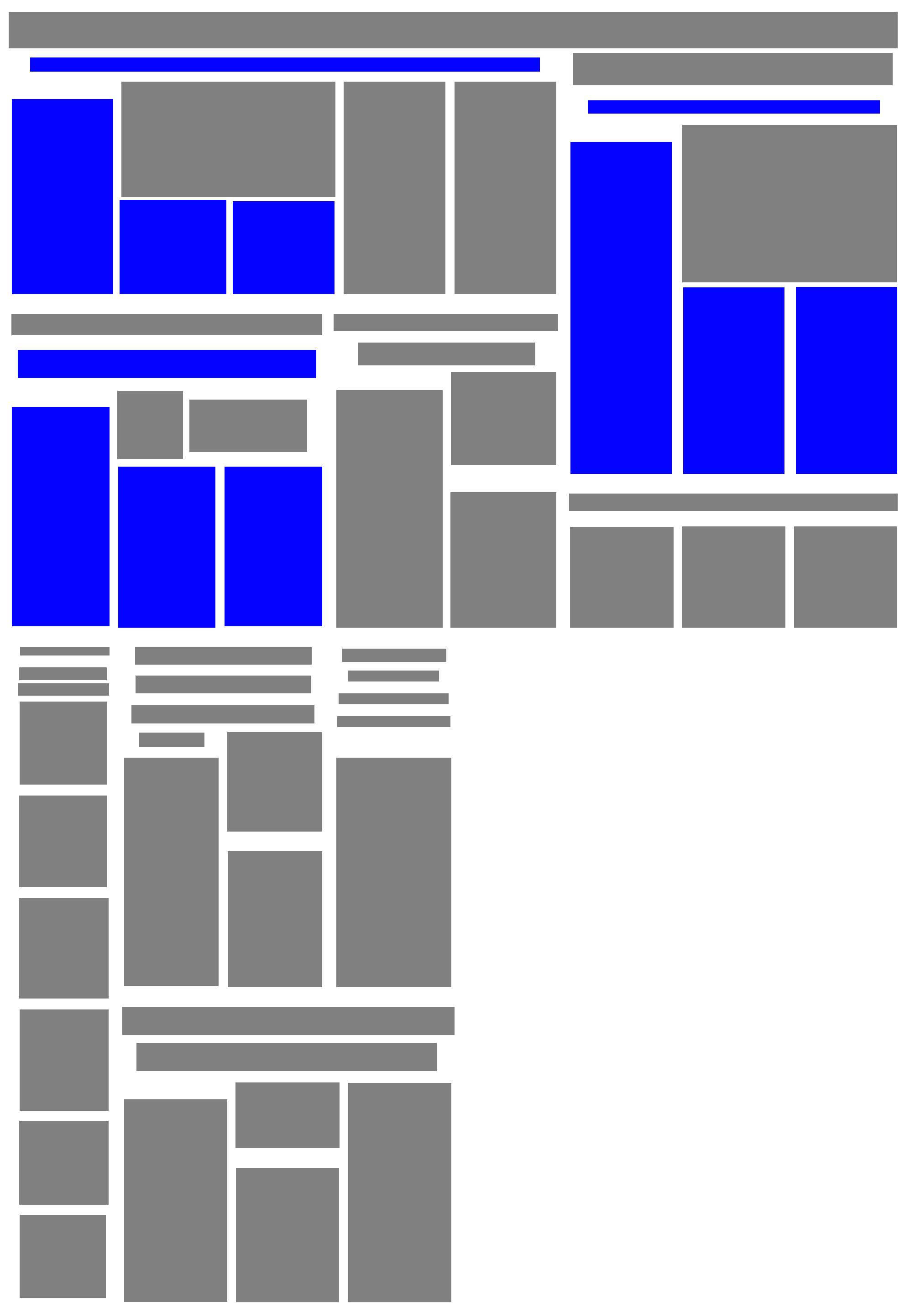}}
\label{R_partial2}} \hspace{0.5mm}\\
\subfloat[]{\fbox{\includegraphics[width = 2.5cm, height = 3.5cm]{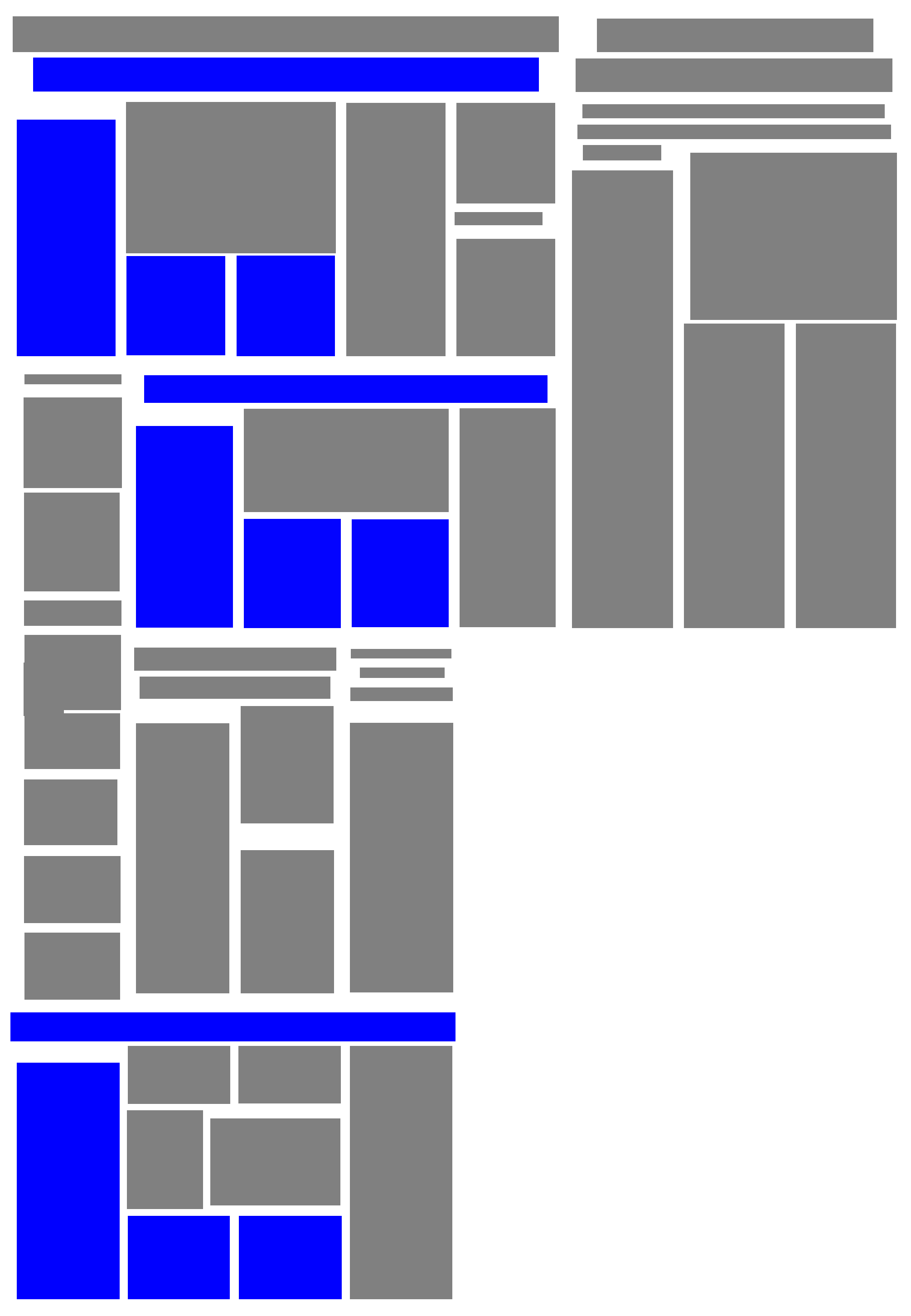}}
\label{R_partial3}} \hspace{0.5mm}
\subfloat[]{\fbox{\includegraphics[width = 2.5cm, height = 3.5cm]{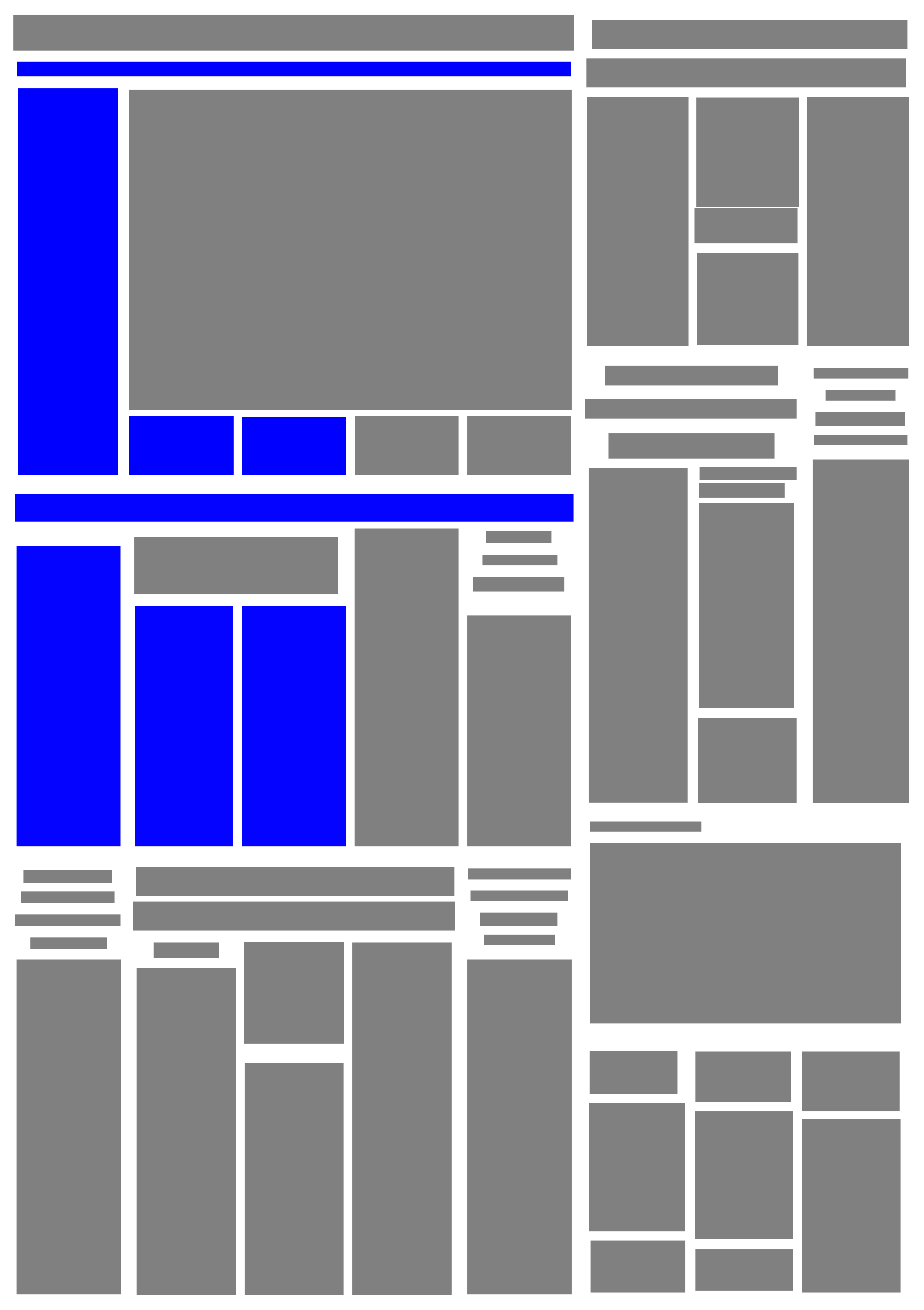}}
\label{R_partial4}} \hspace{0.5mm}
\subfloat[]{\fbox{\includegraphics[width = 2.5cm, height = 3.5cm]{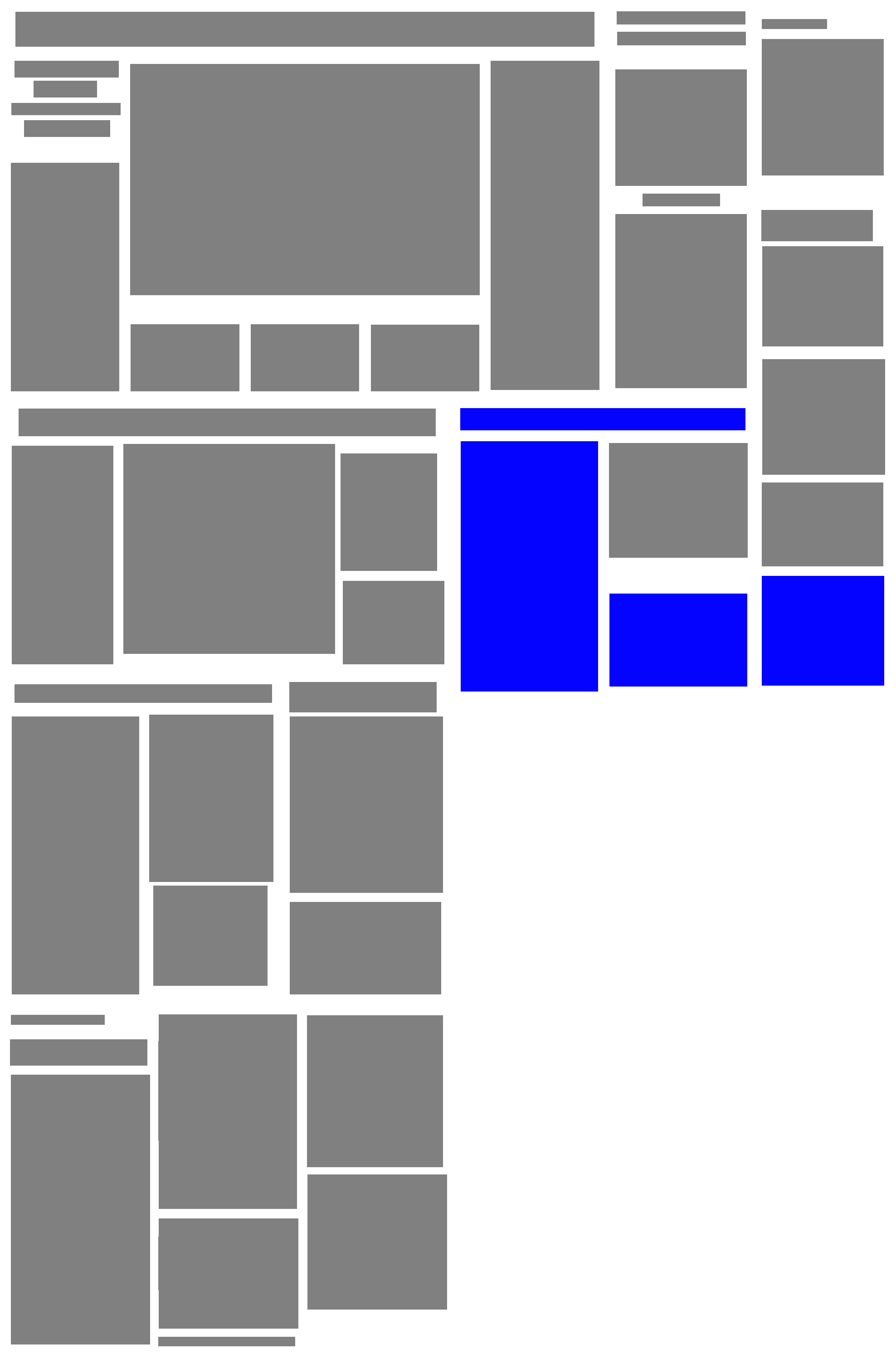}}
\label{R_partial5}}
\caption{Ranked results for the query layout in (a): The system
first generates a dummy block (Sec.~\ref{sec:Searching_and_Retrieval}).
The system ranks results in order of relative average discrepancy
in block aspect ratios, and relative positions
(Sec.~\ref{sec:Searching_and_Retrieval}.)}
\label{fig:Results_Ranked}
\end{figure*}

Fig. \ref{fig:Results_Ranked} shows an example of ranked results
for a Type 4 query layout (Fig.~\ref{R_Partial_Query}).
As mentioned earlier (Sec.~\ref{sec:Searching_and_Retrieval}),
we create a dummy block (Fig.~\ref{R_Partial_Query_Dummy}) 
in the vacant space, which corresponds to one or more missing blocks.
The matched layouts are ranked on the basis of relative average discrepancy
in block aspect ratios, and relative positions, compared to the query blocks.
Table~\ref{table_results} lists
recall and precision statistics on the database of 4776 images, for 
various types of query layouts. 
Recall and precision values are
calculated on the basis of the number of dcuments retrieved  
(`\# Documents'), and not the total number of layouts.
(A document may contain more than one instance of a particular
type of query sub layout, as in Fig.~\ref{fig:Results_Ranked}c.)
\begin{table}[!t]
\caption{Statistics of search and retrieval procedure}
\label{table_results}
\centering
\begin{tabular}{p{1.5cm} p{2.00cm} p{1cm} p{1cm} p{1cm} }
\hline
Query & \# Documents & Recall (\%)& Precision (\%) &
Time(sec)\\
\hline
Type 1 & 1592  & 95.41 & 98.14 & 0.685 \\
\hline
Type 2 & 1986 & 95.21 & 97.91  & 0.696 \\
\hline
Type 3 & 1986 & 95.21 & 97.91  & 0.696 \\
\hline
Type 4 & 2990 & 91.42 & 98.387 & 0.742 \\
\hline
Type 5 & 2990 & 91.42 & 98.387 & 0.742 \\
\hline
Type 6 & 2990 & 91.42 & 98.387 & 0.742 \\
\hline
\end{tabular}
\end{table}
The departure from perfect
statistics is on account of pre-processing errors (in Symmetry
maximization (Sec.~\ref{subsec:symmetry_maximization}) and
generating multiple segmentation hypotheses
(Sec.~\ref{subsec:multi_seg_hypotheses}).
To establish the utility of generating multiple segmentation 
hypotheses, an experiment omitted this stage completely.
The precision value in this case was same, i.e., 98.1\%,
but recall fell to less than 40\%. Reasons include missing out on
plausible hypotheses, and the presence of small and/or noisy
blocks which unnecessarily disturb the document graph structure
(Sec.~\ref{sec:pre_processing_representation}).

\section{Conclusions}
We propose a system to retrieve user-specified combinations of
sub-layouts, which works for highly unstructured documents such
as newspaper images. The graph-based matching strategy integrates
a hashing-based indexing for fast matching, with
handling cases of pre-processing segmentation errors.
We show encouraging retrieval results for a representative sample
database of 4776 newspaper images.

\bibliographystyle{spbasic}
\bibliography{ref}

%

\end{document}